\documentclass[aps,10pt,pre,twocolumn]{revtex4-1}
\usepackage{amsmath,amssymb,amsfonts,amsthm}
\usepackage[ascii]{inputenc}
\usepackage{bbm}
\usepackage{array}
\usepackage{graphicx}
\usepackage{tikz}
\usetikzlibrary{positioning}
\usepackage[caption=false]{subfig}
\usepackage[pdftex,bookmarks=false,colorlinks=true,linkcolor=blue,
citecolor=blue,filecolor=black,urlcolor=blue]{hyperref}

\providecommand{\ud}{\mathrm{d}}
\providecommand{\ptype}[1]{\mathsf{#1}}
\providecommand{\abs}[1]{\left\lvert#1\right\rvert}
\providecommand{\bra}[1]{\left\langle#1\right\vert}
\providecommand{\ket}[1]{\left\vert#1\right\rangle}

\begin{document}

\title{Quantum Boltzmann equation for spin-dependent reactions in the kinetic regime}

\author{Martin L.R. F\"urst}
\email{mfuerst@ma.tum.de}
\affiliation{Excellence Cluster Universe,
Boltzmannstra{\ss}e 2}
\affiliation{Zentrum Mathematik,
Boltzmannstra{\ss}e 3,
Technische Universit\"at M\"unchen,
85747 Garching bei M\"unchen,
Germany}

\author{Markus Kotulla}
\email{markus.kotulla@tum.de}
\affiliation{Physik Department,
James-Franck-Stra{\ss}e 1,
Technische Universit\"at M\"unchen,
85747 Garching bei M\"unchen,
Germany}

\author{Christian B. Mendl}
\email{mendl@ma.tum.de}
\affiliation{Zentrum Mathematik,
Boltzmannstra{\ss}e 3,
Technische Universit\"at M\"unchen,
85747 Garching bei M\"unchen,
Germany}

\author{Herbert Spohn}
\email{spohn@ma.tum.de}
\affiliation{Zentrum Mathematik,
Boltzmannstra{\ss}e 3,
Technische Universit\"at M\"unchen}
\affiliation{Physik Department,
James-Franck-Stra{\ss}e 1,
Technische Universit\"at M\"unchen,
85747 Garching bei M\"unchen,
Germany}

\date{December 19, 2014}

\begin{abstract}
\noindent We derive and analyze an effective quantum Boltzmann equation in the kinetic regime for the interactions of four distinguishable types of fermionic spin-$\frac{1}{2}$ particles, starting from a general quantum field Hamiltonian. Each particle type is described by a time-dependent, $2 \times 2$ spin-density (``Wigner'') matrix. We show that density and energy conservation laws as well as the H-theorem hold, and enumerate additional conservation laws depending on the interaction. The conserved quantities characterize the $t \to \infty$ thermal (Fermi-Dirac) equilibrium state. We illustrate the approach to equilibrium by numerical simulations in the isotropic three-dimensional setting.
\end{abstract}

\maketitle

\section{Introduction}

Spin-dependent interactions on the quantum level give rise to a wide range of phenomena, for example, the quantum coherence preserving charge and energy transfer during photosynthesis \cite{CoherencePhotosynthesis2007,PhotosyntheticAlgae2010}, avian navigation of birds \cite{AvianMagnetoreception2008,MagnetoreceptionBirds2009} or quantum transport in condensed matter physics \cite{MagneticTunnellingReview2003, SpinInjectionDetection2010, SpinFilter2013}, and are even investigated in astrophysics \cite{WIMP2009}. The dynamics can typically be modeled by a Hamiltonian on the level of quantum field theory, but solving the resulting equations is often difficult in practice, such that effective approximations are desirable.

Here, we consider the limit of a weak potential interaction term $\lambda V$ with $\lambda \ll 1$ in a general quantum field Hamiltonian (see Sec.~\ref{sec:Hamiltonian}), and systematically derive and analyze an effective quantum Boltzmann equation in the kinetic regime (Sec.~\ref{sec:Boltzmann}) which describes the interactions of four fermionic spin-$\frac{1}{2}$ particles. In particular, we prove the H-theorem and discuss the conservation laws depending on the interaction (see Sec.~\ref{sec:Properties}), and present a detailed analysis of the relation between the conserved quantities and the $t \to \infty$ thermal equilibrium state (see Sec.~\ref{sec:Stationary}). Finally, we illustrate the approach to equilibrium by numerical simulations in the isotropic three-dimensional setting (Sec.~\ref{sec:Numerics} and \ref{sec:Simulation}). The main differences compared to previous work \cite{BoltzmannFermi2012, BoltzmannNonintegrable2013} are the four particle types and the continuous domain for the momentum.

\section{Multi-component field Hamiltonian}
\label{sec:Hamiltonian}

We consider fermionic spin-$\frac{1}{2}$ fields in a $d$-dimensional box $U = [-\ell, \ell]^{d}$, with creation and annihilation operators $a^{\alpha}_{\sigma}(x)^*$, $a^{\alpha}_{\sigma}(x)$, where $\sigma \in \{\uparrow , \downarrow \}$ denotes the spin and $\alpha \in \{\ptype{a}, \ptype{b}, \ptype{c}, \ptype{d}\}$ the particle type. The operators for the same type obey the fermionic anticommutator relations
\begin{equation}
\begin{split}
\left\{\, a^{\alpha}_{\sigma}(x)^*,a^{\alpha}_{\tau}(y) \hspace{5pt} \right\} &= \delta(x-y) \, \delta_{\sigma\tau}, \\
\left\{\, a^{\alpha}_{\sigma}(x), \hspace{4pt} a^{\alpha}_{\tau}(y) \hspace{5pt} \right\} &= 0, \\
\left\{\, a^{\alpha}_{\sigma}(x)^*,a^{\alpha}_{\tau}(y)^*  \right\} &= 0
\end{split}
\end{equation}
with $\{A, B\} = A B + B A$. The operators for differing particles commute, i.e.,
\begin{equation}
[ a^{\alpha}_{\sigma}(x)^*, a^{\beta}_{\tau}(y) ] = 0, \quad \alpha \neq \beta.
\end{equation}
with the commutator $[A, B] = A B - B A$. 

Formally, the underlying one-particle Hilbert space for each particle type is $L^2(U, \mathbb{C}^2)$, and the full Hilbert space is the tensor product of the Fock spaces for the individual particle types. 

Our field Hamiltonian is given by
\begin{equation}
\label{eq:Hamiltonian}
H = H_0 + \lambda H_1
\end{equation}
with $0 < \lambda \ll 1$ and
\begin{equation}
\label{eq:H0defSpatial}
H_0 = \int_{U} \ud x\, a(x)^* \cdot \hat{\omega}(x) \cdot a(x)
\end{equation}
as well as
\begin{equation}
\label{eq:H1defSpatial}
\begin{split}
H_1 &= \int_{U^4} \ud x_{1234} \\
&\Big[ \ \big(a(x_1)^* \cdot \mathbf{V}^\ptype{ab} \cdot a(x_2)\big) \big(a(x_3)^{*} \cdot \mathbf{V}^\ptype{cd} \cdot a(x_4)\big) \\
& + \big(a(x_1)^{*} \cdot \mathbf{V}^\ptype{ad} \cdot a(x_2)\big) \big(a(x_3)^{*} \cdot \mathbf{V}^\ptype{cb} \cdot a(x_4)\big) \Big] + \mathrm{h.c.}
\end{split}
\end{equation}
Here, the $a(x)$ are operator-valued vectors
\begin{equation}
a(x) = \big( a^{\ptype{a}}_{\uparrow}(x), a^{\ptype{a}}_{\downarrow}(x), a^{\ptype{b}}_{\uparrow}(x), \dots, a^{\ptype{d}}_{\downarrow}(x) \big)
\end{equation}
and $\mathbf{V}^{\alpha\beta}$ are $8 \times 8$ matrices to be specified below ($\alpha, \beta \in \{ \ptype{a}, \ptype{b}, \ptype{c}, \ptype{d} \}$). $\hat{\omega}(x)$ in Eq.~\eqref{eq:H0defSpatial} is the Fourier transform of the dispersion relation.

Historically, Enrico Fermi derived \cite{Fermi1934} an explanation of the $\beta$ decay using a Hamiltonian of the form \eqref{eq:Hamiltonian}. Fermi's four-fermion theory could also predict the weak interaction remarkably well. In this work, our aim is a generalization to spin-dependent interactions.

We will use the following convention for the Fourier transform (corresponding to the finite volume $U$)
\begin{equation}
\hat{f}(p) = \int_{U} \ud x \, f(x) \,\mathrm{e}^{-\mathrm{i}\,p \cdot x}
\end{equation}
and the inverse Fourier transform
\begin{equation}
f(x) = \frac{1}{|U|}\sum_{p \in \hat{U}}\hat{f}(p)\, \mathrm{e}^{\mathrm{i}\,p \cdot x},
\end{equation}
with $L = 2\ell$ and $\hat{U} = \frac{2\pi}{L} \mathbb{Z}^d$. $|U| = L^d$ denotes the volume of the box. Accordingly, the anticommutator relations in momentum space read
\begin{equation}
\begin{split}
\left\{\, \hat{a}^{\alpha}_{\sigma}(p)^*,\hat{a}^{\alpha}_{\tau}(p') \hspace{6pt} \right\} &= \abs{U}\, \delta(p-p')\, \delta_{\sigma\tau}, \\
\left\{\, \hat{a}^{\alpha}_{\sigma}(p), \hspace{4pt} \hat{a}^{\alpha}_{\tau}(p') \hspace{6pt} \right\} &= 0, \\
\left\{\, \hat{a}^{\alpha}_{\sigma}(p)^*,\hat{a}^{\alpha}_{\tau}(p')^* \right\} &= 0.
\end{split}
\end{equation}


The kinetic part of the Hamiltonian in momentum space reads
\begin{equation}
\label{eq:H0def}
H_0 = \frac{1}{\abs{U}} \sum_{p \in \hat{U}} \hat{a}(p)^* \cdot \omega(p) \cdot \hat{a}(p).
\end{equation}
Here, the dispersion relations
\begin{equation}
\label{eq:omegaAlphaDef}
\omega^\alpha(p) = \frac{\abs{p}^2}{2 m^\alpha}
\end{equation}
for each particle $\alpha$ with mass $m^{\alpha}$ are summarized in the $8 \times 8$ diagonal matrix
\begin{equation}
\label{eq:omegaDef}
\omega(p) = \mathrm{diag}\!\left[ \omega^{\ptype{a}}(p), \omega^{\ptype{b}}(p), \omega^{\ptype{c}}(p), \omega^{\ptype{d}}(p) \right] \otimes \mathbbm{1}_{2 \times 2}.
\end{equation}
The $2 \times 2$ identity matrices appear in spin space since the kinetic energy is independent of spin. The \emph{finite} box $U \subset \mathbb{R}^d$ ensures that the Fourier transform of the dispersion relation in Eq.~\eqref{eq:omegaAlphaDef} is well-defined.

The interaction part of the Hamiltonian in momentum space is given by
\begin{equation}
\label{eq:H1def}
H_1 = H_1^{\ptype{a}\ptype{b}\ptype{c}\ptype{d}} + H_1^{\ptype{a}\ptype{d}\ptype{c}\ptype{b}} + \mathrm{h.c.}
\end{equation}
with
\begin{equation}
\big( H_1^{\alpha\beta\gamma\delta} \big)^* = H_1^{\delta\gamma\beta\alpha}
\end{equation}
and
\begin{equation}
\label{eq:H1abcd_def}
\begin{split}
H_1^{\alpha\beta\gamma\delta} &= \frac{1}{\abs{U}^{4}} \sum_{p_{1234}} \delta(\underline{p}) \\
& \big(\hat{a}(p_1)^* \cdot \mathbf{V}^{\alpha\beta} \cdot \hat{a}(p_2)\big) \big(\hat{a}(p_3)^{*} \cdot \mathbf{V}^{\gamma\delta} \cdot \hat{a}(p_4)\big).
\end{split}
\end{equation}
Here $\underline{p} = p_1 - p_2 + p_3 - p_4$ is the momentum difference, $\sum_{p_{1234}} = \sum_{p_1, p_2, p_3, p_4}$, and we have introduced the 8$\times$8 matrices
\begin{equation}
\label{eq:VabVcd}
\mathbf{V}^\ptype{ab} =
\begin{pmatrix}
0 & V^\ptype{ab} & 0 & 0\\
0 & 0 & 0 & 0\\
0 & 0 & 0 & 0\\
0 & 0 & 0 & 0
\end{pmatrix}, \quad
\mathbf{V}^\ptype{cd} =
\begin{pmatrix}
0 & 0 & 0 & 0\\
0 & 0 & 0 & 0\\
0 & 0 & 0 & V^\ptype{cd}\\
0 & 0 & 0 & 0
\end{pmatrix},
\end{equation}
\begin{equation}
\label{eq:VadVcb}
\mathbf{V}^\ptype{ad} =
\begin{pmatrix}
0 & 0 & 0 & V^\ptype{ad}\\
0 & 0 & 0 & 0\\
0 & 0 & 0 & 0\\
0 & 0 & 0 & 0
\end{pmatrix}, \quad
\mathbf{V}^\ptype{cb} =
\begin{pmatrix}
0 & 0 & 0 & 0\\
0 & 0 & 0 & 0\\
0 & V^\ptype{cb} & 0 & 0\\
0 & 0 & 0 & 0
\end{pmatrix}.
\end{equation}

The Hamiltonian should model the interactions
\begin{equation}
\label{eq:interact_diag}
\begin{tikzpicture}[baseline=(current bounding box.center)]
\node                   (a) {$\ptype{a}$};
\node[right of=a]       (b) {$\ptype{b}$};
\node[below=1.5pt of a] (c) {$\ptype{c}$};
\node[right of=c]       (d) {$\ptype{d}$};
\draw[<-] (a) -- (b);
\draw[<-] (c) -- (d);
\end{tikzpicture},
\qquad
\begin{tikzpicture}[baseline=(current bounding box.center)]
\node                   (a) {$\ptype{a}$};
\node[right of=a]       (b) {$\ptype{b}$};
\node[below=1.5pt of a] (c) {$\ptype{c}$};
\node[right of=c]       (d) {$\ptype{d}$};
\draw[->] (a) -- (b);
\draw[->] (c) -- (d);
\end{tikzpicture}
\end{equation}
and
\begin{equation}
\label{eq:interact_cross}
\begin{tikzpicture}[baseline=(current bounding box.center)]
\node                   (a) {$\ptype{a}$};
\node[right of=a]       (b) {$\ptype{b}$};
\node[below=1.5pt of a] (c) {$\ptype{c}$};
\node[right of=c]       (d) {$\ptype{d}$};
\draw[<-] (a) -- (d);
\draw[<-] (c) -- (b);
\end{tikzpicture},
\qquad
\begin{tikzpicture}[baseline=(current bounding box.center)]
\node                   (a) {$\ptype{a}$};
\node[right of=a]       (b) {$\ptype{b}$};
\node[below=1.5pt of a] (c) {$\ptype{c}$};
\node[right of=c]       (d) {$\ptype{d}$};
\draw[->] (a) -- (d);
\draw[->] (c) -- (b);
\end{tikzpicture}
\end{equation}
To quantify the (possibly spin-dependent) strength of the interactions, we introduce the $2 \times 2$ real-valued ``interaction matrices'' $V^\ptype{ab}$, $V^\ptype{cd}$, $V^\ptype{ad}$ and $V^\ptype{cb}$ in momentum space. They model the interactions
\begin{equation}
V^{\alpha\beta}: \alpha \longleftarrow \beta, \qquad (V^{\alpha \beta})^* = V^{\beta \alpha}: \alpha \longrightarrow \beta
\end{equation}
with $\alpha, \beta \in \{ \ptype{a}, \ptype{b}, \ptype{c}, \ptype{d} \}$. For simplicity, we assume that these matrices are constant (independent of $p$). Note that they permit spin dependent reactions like
\begin{equation}
(\ptype{a}_\uparrow, \ptype{c}_\downarrow) \longrightarrow (\ptype{b}_\downarrow, \ptype{d}_\downarrow).
\end{equation}

\medskip

The system respects conservation of energy and overall particle number. We denote the particle number operator for field $\alpha$ by
\begin{equation}
\hat{N}^{\alpha} = \sum_{p \in \hat{U}, \sigma \in \{\uparrow, \downarrow\}}\hat{a}_{\sigma}^{\alpha}(p)^* \, \hat{a}_{\sigma}^{\alpha}(p)
\end{equation}
and thus the total particle number operator reads
\begin{equation}
\hat{N} = \sum_{\alpha \in \{\ptype{a}, \ptype{b}, \ptype{c}, \ptype{d}\}} \hat{N}^{\alpha}.
\end{equation}
It satisfies the relation $[H, \hat{N}] = 0$, as required. Certain sums of two particles are also conserved,
\begin{equation}
\label{eq:HamiltonParticleCons}
\begin{split}
[H, \hat{N}^\ptype{a} + \hat{N}^\ptype{b}] &= 0, \quad [H, \hat{N}^\ptype{c} + \hat{N}^\ptype{d}] = 0,\\
[H, \hat{N}^\ptype{a} + \hat{N}^\ptype{d}] &= 0, \quad [H, \hat{N}^\ptype{c} + \hat{N}^\ptype{b}] = 0,\\
\end{split}
\end{equation}
since the Hamiltonian only includes the processes in Eq.~\eqref{eq:interact_diag} and \eqref{eq:interact_cross}. Concerning $\hat{N}^\ptype{a} + \hat{N}^\ptype{b}$, for example, the creation of $\ptype{a}$ involves a simultaneous annihilation of $\ptype{b}$ according to the Hamiltonian structure \eqref{eq:H1abcd_def} and hence the sum $\hat{N}^\ptype{a} + \hat{N}^\ptype{b}$ remains constant. Note that not all combinations of two particle types are conserved, e.g.,
\begin{equation}
[H, \hat{N}^\ptype{a} + \hat{N}^\ptype{c}] \neq 0, \quad [H, \hat{N}^\ptype{b} + \hat{N}^\ptype{d}] \neq 0.
\end{equation}

\section{Boltzmann kinetic equation}
\label{sec:Boltzmann}

We will derive the kinetic Boltzmann equation in appendix~\ref{sec:Derivation}. The central object are the two-point functions $W^{\alpha}(p,t)$, defined by the relation
\begin{equation}
\langle \hat{a}^\alpha_\sigma(p,t)^* \, \hat{a}^\beta_\tau(p',t) \rangle = |U| \delta_{\alpha\beta} \, \delta(p-p') W^\alpha_{\sigma\tau}(p,t)
\end{equation}
for all particle types $\alpha, \beta \in \{ \ptype{a}, \ptype{b}, \ptype{c}, \ptype{d} \}$. We collect the $2\times2$ positive semidefinite (spin density) Wigner states $W^\alpha(p,t)$ in a $8 \times 8$ block-diagonal matrix,
\begin{equation}
W_1 = \mathrm{diag}\left[W^{\ptype{a}}_1, W^{\ptype{b}}_1, W^{\ptype{c}}_1, W^{\ptype{d}}_1\right],
\end{equation}
where we have used the notation $W_1 = W(p_1,t)$. The resulting Boltzmann equation reads
\begin{equation}
\label{eq:BoltzmannEquation}
\frac{\partial}{\partial t} W(p,t)
= \mathcal{C}[W](p,t)
\end{equation}
with the collision operator consisting of a conservative and dissipative part,
\begin{equation}
\mathcal{C}[W](p,t) = \mathcal{C}_\mathrm{cons}[W](p,t) + \mathcal{C}_{\mathrm{diss}}[W](p,t).
\end{equation}
$\mathcal{C}_\mathrm{cons}$ and $\mathcal{C}_{\mathrm{diss}}$ both preserve the block-diagonal structure.

%
The conservative collision operator $\mathcal{C}_{\mathrm{cons}}$ is the Vlasov-type operator
\begin{equation}
\label{eq:Cc}
\mathcal{C}_\mathrm{cons}[W](p,t) = -\mathrm{i}\,[H_\mathrm{eff}(p,t), W(p,t)],
\end{equation}
where the effective Hamiltonian $H_\mathrm{eff}(p,t)$ is a $8 \times 8$ block-diagonal matrix which itself depends on $W$:
\begin{equation}
\label{eq:Heff}
H_{\mathrm{eff},1} = \frac{1}{(2\pi)^3} \int_{\mathbb{R}^{3d}} \ud p_{234} \, \delta(\underline{p}) \, \mathcal{P}\big(\underline{\omega}^{-1}\big) \, h_{\mathrm{eff}}[W]_{234}.
\end{equation}
The energy differences are defined as
\begin{equation}
\label{eq:generalizedenergies}
\underline{\omega} = \mathrm{diag}\!\left[\omega^\ptype{abcd}, \omega^\ptype{badc}, \omega^\ptype{cdab}, \omega^\ptype{dcba}\right] \otimes \mathbbm{1}_{2\times2}
\end{equation}
with $\omega^{\alpha\beta\gamma\delta} = \omega^{\alpha}(p_1) - \omega^{\beta}(p_2) + \omega^{\gamma}(p_3) - \omega^{\delta}(p_4)$. In Eq.~\eqref{eq:Heff} we have used the shorthand notation $\ud p_{234} = \ud p_2\,\ud p_3\,\ud p_4$. Note that the expression $\mathcal{P}(\underline{\omega}^{-1})$ is a diagonal \emph{matrix} of principal values. The index $234$ means that the block-diagonal matrix $h_{\mathrm{eff}}[W]$ depends on $p_2$, $p_3$, and $p_4$. It is given by
\begin{equation}
\label{eq:heff}
\begin{split}
&h_{\mathrm{eff}}[W]_{234} = \\
&\quad - V^{\scriptscriptstyle{=}}\, W_2\, V^{\mathsf{x}}\, \tilde{W}_3\, V^{\scriptscriptstyle{=}}\, W_4\, V^{\mathsf{x}} - V^{\mathsf{x}}\, W_4\, V^{\scriptscriptstyle{=}}\, \tilde{W}_3\, V^{\mathsf{x}}\, W_2\, V^{\scriptscriptstyle{=}} \\
&\quad - V^{\scriptscriptstyle{=}}\, \tilde{W}_2\, V^{\mathsf{x}}\, W_3\, V^{\scriptscriptstyle{=}}\, \tilde{W}_4\, V^{\mathsf{x}} - V^{\mathsf{x}}\, \tilde{W}_4\, V^{\scriptscriptstyle{=}}\, W_3\, V^{\mathsf{x}}\, \tilde{W}_2\, V^{\scriptscriptstyle{=}} \\
&\quad - V^{\scriptscriptstyle{=}}\, W_2\, V^{\scriptscriptstyle{=}} \cdot \mathfrak{tr}\big[Y\, \tilde{W}_3\, V^{\scriptscriptstyle{=}}\, W_4\, V^{\scriptscriptstyle{=}}\, Y\big] \\
&\quad - V^{\scriptscriptstyle{=}}\, \tilde{W}_2\, V^{\scriptscriptstyle{=}} \cdot \mathfrak{tr}\big[Y\, W_3\, V^{\scriptscriptstyle{=}}\, \tilde{W}_4\, V^{\scriptscriptstyle{=}}\, Y\big] \\
&\quad - V^{\mathsf{x}}\, W_4\, V^{\mathsf{x}}\; \cdot \mathfrak{tr}\big[Y\, \tilde{W}_3\, V^{\mathsf{x}}\;\, W_2\, V^{\mathsf{x}}\, Y\big] \\
&\quad - V^{\mathsf{x}}\, \tilde{W}_4\, V^{\mathsf{x}}\; \cdot \mathfrak{tr}\big[Y\, W_3\, V^{\mathsf{x}}\;\, \tilde{W}_2\, V^{\mathsf{x}}\, Y\big],
\end{split}
\end{equation}
using the notation $\tilde{W}_i = \mathbbm{1}_{8 \times 8} - W_i$. The $\mathfrak{tr}$ operator appearing in Eq.~\eqref{eq:heff} acts separately on each $(2 \times 2)$ diagonal block, i.e.,
\begin{equation}
\mathfrak{tr}[W] = \sum_{\alpha \in \{\ptype{a}, \ptype{b}, \ptype{c}, \ptype{d}\}} E^{\alpha} \, \mathrm{tr}[E^{\alpha} \, W], \quad E^{\alpha} = \ket{e_{\alpha}}\bra{e_{\alpha}} \otimes \mathbbm{1}_{2 \times 2}
\end{equation}
with $e_{\alpha}$ enumerating the standard basis of $\mathbb{R}^4$.
The operator $Y$ appearing in Eq.~\eqref{eq:heff} switches the particle types $(\ptype{a},\ptype{b}) \leftrightarrow (\ptype{c},\ptype{d})$ and is defined as
\begin{equation}
Y =
\begin{pmatrix}
0 & 0 & 1 & 0 \\
0 & 0 & 0 & 1 \\
1 & 0 & 0 & 0 \\
0 & 1 & 0 & 0
\end{pmatrix}
\otimes \mathbbm{1}_{2 \times 2}.
\end{equation}
The $8 \times 8$ interaction matrices read
\begin{equation}
\label{eq:VdiagDef}
V^{\scriptscriptstyle{=}} =
\begin{pmatrix}
0 & V^\ptype{ab} & 0 & 0\\
V^\ptype{ba} & 0 & 0 & 0\\
0 & 0 & 0 & V^\ptype{cd}\\
0 & 0 & V^\ptype{dc} & 0
\end{pmatrix}
\end{equation}
and
\begin{equation}
\label{eq:VcrossDef}
V^{\mathsf{x}} =
\begin{pmatrix}
0 & 0 & 0 & V^\ptype{ad}\\
0 & 0 & V^\ptype{bc} & 0\\
0 & V^\ptype{cb} & 0 & 0\\
V^\ptype{da} & 0 & 0 & 0
\end{pmatrix},
\end{equation}
where always $V^{\beta \alpha} = (V^{\alpha \beta})^*$. The superscripts of $V^{\scriptscriptstyle{=}}$ and $V^{\mathsf{x}}$ refer to the arrows in Eq.~\eqref{eq:interact_diag} and \eqref{eq:interact_cross}.

It turns out that the interaction matrices enter the collision operator only via the following $4 \times 4$ matrix,
\begin{equation}
\label{eq:Vop}
\mathcal{V} = \left(V^\ptype{ab} \otimes V^\ptype{cd}\right) + \left(V^\ptype{ad} \otimes V^\ptype{cb}\right) T,
\end{equation}
with
\begin{equation}
T = \begin{pmatrix}
1 &   &   &   \\
  & 0 & 1 &   \\
  & 1 & 0 &   \\
  &   &   & 1
\end{pmatrix} \in \mathbbm{R}^{4 \times 4}
\end{equation}
an operator which interchanges tensor components (represented in the standard basis $\ket{\uparrow\uparrow}$, $\ket{\uparrow\downarrow}$, $\ket{\downarrow\uparrow}$, $\ket{\downarrow\downarrow}$). For example, the $(\ptype{a},\ptype{a})$-component (first $2 \times 2$ block) of the integrand $h_{\mathrm{eff}}[W]_{234}$ can be represented as
\begin{equation}
\label{eq:heff_aa}
\begin{split}
&\bra{\sigma} \big( h_{\mathrm{eff}}[W]_{234} \big)^{\ptype{a}} \ket{\tau} \\
&= \mathrm{tr}\Big[ \big( \ket{\tau}\bra{\sigma} \otimes \tilde{W}^\ptype{c}_3 \big) \, \mathcal{V} \, \big(W^\ptype{b}_2 \otimes W^\ptype{d}_4\big) \, \mathcal{V}^* \\
&\hspace{13pt} + \big( \ket{\tau}\bra{\sigma} \otimes W^\ptype{c}_3 \big) \, \mathcal{V} \, \big(\tilde{W}^\ptype{b}_2 \otimes \tilde{W}^\ptype{d}_4\big) \, \mathcal{V}^*
\Big]
\end{split}
\end{equation}
with the notation $\tilde{W}^{\alpha}_i = \mathbbm{1}_{2 \times 2} - W^{\alpha}_i$. Note that $h_{\mathrm{eff}}[W]_{234}$ is invariant under $W^{\alpha}_i \leftrightarrow \tilde{W}^{\alpha}_i$, and formally similar to Eq.~\eqref{eq:AATa_tensor}. The other components arise from the $(\ptype{a},\ptype{a})$-component by permutations of $\ptype{a}$, $\ptype{b}$, $\ptype{c}$, $\ptype{d}$, as for the dissipative operator.

\medskip

The dissipative part of the collision operator is
\begin{equation}
\label{eq:Cd}
\begin{split}
\mathcal{C}_{\mathrm{diss}}[W]_1 &= \frac{\pi}{(2\pi)^3} \int_{\mathbb{R}^{3d}} \ud p_{234} \, \delta(\underline{p}) \, \delta(\underline{\omega}) \\
&\qquad\quad \cdot \big( \mathcal{A}_{\mathrm{quad}}[W]_{1234} + \mathcal{A}_{\mathrm{tr}}[W]_{1234} \big),
\end{split}
\end{equation}
where the index $1234$ means that the block-diagonal matrices $\mathcal{A}_{\mathrm{quad}}[W]$ and $\mathcal{A}_{\mathrm{tr}}[W]$ depend on $p_1$, $p_2$, $p_3$, and $p_4$.
%
They are given by
\begin{equation}
\label{eq:Aquad}
\begin{split}
&\mathcal{A}_{\mathrm{quad}}[W]_{1234} = \\
&+ \tilde{W}_1\, V^{\scriptscriptstyle{=}}\, W_2\, V^{\mathsf{x}}\, \tilde{W}_3\, V^{\scriptscriptstyle{=}}\, W_4\, V^{\mathsf{x}} \\
&- W_1\, V^{\scriptscriptstyle{=}}\, \tilde{W}_2\, V^{\mathsf{x}}\, W_3\, V^{\scriptscriptstyle{=}}\, \tilde{W}_4\, V^{\mathsf{x}} \\
&+ \tilde{W}_1\, V^{\mathsf{x}}\, W_4\, V^{\scriptscriptstyle{=}}\, \tilde{W}_3\, V^{\mathsf{x}}\, W_2\, V^{\scriptscriptstyle{=}} \\
&- W_1\, V^{\mathsf{x}}\, \tilde{W}_4\, V^{\scriptscriptstyle{=}}\, W_3\, V^{\mathsf{x}}\, \tilde{W}_2\, V^{\scriptscriptstyle{=}} + \mathrm{h.c.}
\end{split}
\end{equation}
and
\begin{equation}
\label{eq:Atr}
\begin{split}
&\mathcal{A}_{\mathrm{tr}}[W]_{1234} = \\
&+ \big(\tilde{W}_1\, V^{\scriptscriptstyle{=}}\, W_2\, V^{\scriptscriptstyle{=}} + \mathrm{h.c.}\big) \cdot \mathfrak{tr}\big[Y\, \tilde{W}_3\, V^{\scriptscriptstyle{=}}\, W_4\, V^{\scriptscriptstyle{=}}\, Y\big] \\
&- \big(W_1\, V^{\scriptscriptstyle{=}}\, \tilde{W}_2\, V^{\scriptscriptstyle{=}} + \mathrm{h.c.}\big) \cdot \mathfrak{tr}\big[Y\, W_3\, V^{\scriptscriptstyle{=}}\, \tilde{W}_4\, V^{\scriptscriptstyle{=}}\, Y\big] \\
&+ \big(\tilde{W}_1\, V^{\mathsf{x}}\, W_4\, V^{\mathsf{x}} \hspace{4pt} + \mathrm{h.c.}\big) \cdot \mathfrak{tr}\big[Y\, \tilde{W}_3\, V^{\mathsf{x}}\, W_2\, V^{\mathsf{x}}\, Y\big] \\
&- \big(W_1\, V^{\mathsf{x}}\, \tilde{W}_4\, V^{\mathsf{x}} \hspace{4pt} + \mathrm{h.c.}\big) \cdot \mathfrak{tr}\big[Y\, W_3\, V^{\mathsf{x}}\, \tilde{W}_2\, V^{\mathsf{x}}\, Y\big].
\end{split}
\end{equation}
If any of the two matrices $V^{\scriptscriptstyle{=}}$ or $V^{\mathsf{x}}$ is zero, then $\mathcal{A}_{\mathrm{quad}}[W] = 0$, and the first two or last two terms of $\mathcal{A}_{\mathrm{tr}}[W]$ disappear.
Note that $W^{\alpha}_i \leftrightarrow \tilde{W}^{\alpha}_i$ effectively switches signs in Eqs.~\eqref{eq:Aquad} and \eqref{eq:Atr}, and that the respective last two terms equal the first two after switching $\ptype{b} \leftrightarrow \ptype{d}$ and $2 \leftrightarrow 4$.

Performing the matrix multiplications in Eq.~\eqref{eq:Aquad} and \eqref{eq:Atr} shows that Wigner matrices with particle types $\alpha$ and $\beta$ are always coupled by the respective $V^{\alpha \beta}$ matrix, e.g., $\tilde{W}^{\alpha}_i \, V^{\alpha \beta} \, W^{\beta}_j$. Additionally, the $(\ptype{b},\ptype{b})$-component arises from the $(\ptype{a},\ptype{a})$-component by permuting $\ptype{a} \leftrightarrow \ptype{b}$, $\ptype{c} \leftrightarrow \ptype{d}$. Analogously, the $(\ptype{c},\ptype{c})$-component arises from $(\ptype{a},\ptype{a})$ by permuting $\ptype{a} \leftrightarrow \ptype{c}$, $\ptype{b} \leftrightarrow \ptype{d}$, and the $(\ptype{d},\ptype{d})$-component arises from the $(\ptype{a},\ptype{a})$-component by permuting $\ptype{a} \leftrightarrow \ptype{d}$, $\ptype{b} \leftrightarrow \ptype{c}$.

Algebraic reformulation of the $(\ptype{a},\ptype{a})$-component of the integrand $\mathcal{A}_{\mathrm{quad}}[W]_{1234} + \mathcal{A}_{\mathrm{tr}}[W]_{1234}$ results in
\begin{equation}
\label{eq:AATa_tensor}
\begin{split}
&\bra{\sigma} \big( \mathcal{A}_{\mathrm{quad}}[W]_{1234} + \mathcal{A}_{\mathrm{tr}}[W]_{1234} \big)^{\ptype{a}} \ket{\tau} \\
&= \mathrm{tr}\Big[ \big( \{ \tilde{W}^\ptype{a}_1, \ket{\tau}\bra{\sigma}\} \otimes \tilde{W}^\ptype{c}_3 \big) \, \mathcal{V} \, \big(W^\ptype{b}_2 \otimes W^\ptype{d}_4\big) \, \mathcal{V}^* \\
&\quad \ - \big( \{ W^\ptype{a}_1, \ket{\tau}\bra{\sigma}\} \otimes W^\ptype{c}_3 \big) \, \mathcal{V} \, \big(\tilde{W}^\ptype{b}_2 \otimes \tilde{W}^\ptype{d}_4\big) \, \mathcal{V}^*
\Big]
\end{split}
\end{equation}
for all spin components $\sigma, \tau$, where $\{ \cdot, \cdot \}$ denotes the anticommutator. Equivalent expressions give the $\ptype{b}$, $\ptype{c}$ and $\ptype{d}$ components after appropriate interchanges of $\ptype{a}$, $\ptype{b}$, $\ptype{c}$, $\ptype{d}$ as above, with the anticommutator acting on $W^\ptype{b}$, $W^\ptype{c}$ and $W^\ptype{d}$, respectively. For example, after a short reformulation
\begin{equation}
\label{eq:AATb_tensor}
\begin{split}
&\bra{\sigma} \big( \mathcal{A}_{\mathrm{quad}}[W]_{1234} + \mathcal{A}_{\mathrm{tr}}[W]_{1234} \big)^{\ptype{b}} \ket{\tau} \\
&= \mathrm{tr}\Big[ - \big( \tilde{W}^\ptype{a}_2 \otimes \tilde{W}^\ptype{c}_4 \big) \, \mathcal{V} \, \big(\{ W^\ptype{b}_1, \ket{\tau}\bra{\sigma}\} \otimes W^\ptype{d}_3\big) \, \mathcal{V}^* \\
&\hspace{26pt} + \big( W^\ptype{a}_2 \otimes W^\ptype{c}_4 \big) \, \mathcal{V} \, \big( \{ \tilde{W}^\ptype{b}_1, \ket{\tau}\bra{\sigma}\} \otimes \tilde{W}^\ptype{d}_3\big) \, \mathcal{V}^*
\Big].
\end{split}
\end{equation}

\section{General properties of the kinetic equation}
\label{sec:Properties}

The kinetic equation inherits density and energy conservation laws of the Hamiltonian system, as shown below, and the H-theorem holds. Specifically for the multi-component system, there emerge additional conserved quantities depending on the special structure of the $V^{\alpha \beta}$ matrices. In this context, the evolution dynamics is invariant under unitary rotations with fixed unitary $U^{\alpha} \in \mathrm{SU}(2)$ (separately for each block and independent of $p$ and $t$), i.e., simultaneously
\begin{equation}
\label{eq:unitary_invariance}
\begin{split}
&W^{\alpha} \to U^{\alpha} W^{\alpha} \left(U^{\alpha}\right)^*,\\
&V^{\alpha \beta} \to U^{\alpha} V^{\alpha \beta} \left(U^{\beta}\right)^*,\\
&\mathcal{V} \to \big(U^{\ptype{a}} \otimes U^{\ptype{c}}\big)\, \mathcal{V}\, \big(U^{\ptype{b}} \otimes U^{\ptype{d}}\big)^*,
\end{split}
\end{equation}
which can be seen from the representation in Eq.~\eqref{eq:AATa_tensor}.

\subsection{Density conservation}
\label{sec:DensityConversation}

We define the spin density matrix of particle type $\alpha$ as
\begin{equation}
\label{eq:SpinDensityDef}
\rho^{\alpha}(t) = \int_{\mathbb{R}^d} \ud p\, W^{\alpha}(p,t),
\end{equation}
and the total spin density matrix as
\begin{equation}
\label{eq:TotalSpinDensityDef}
\rho(t) = \sum_{\alpha \in \{\ptype{a}, \ptype{b}, \ptype{c}, \ptype{d}\}} \rho^{\alpha}(t).
\end{equation}

The analogue of the particle conservation $[H, N] = 0$ on the kinetic level reads
\begin{equation}
\label{eq:DensityCons}
\begin{split}
\frac{\ud}{\ud t} \mathrm{tr}[\rho(t)]
&= \frac{\ud}{\ud t} \int_{\mathbb{R}^d} \ud p\, \mathrm{tr}[W(p,t)] \\
&= \int_{\mathbb{R}^d} \ud p\, \mathrm{tr}[\mathcal{C}[W](p,t)] = 0.
\end{split}
\end{equation}
Even more strongly, according to Eq.~\eqref{eq:HamiltonParticleCons} it should hold that
\begin{equation}
\label{eq:DensitySubsetCons}
\frac{\ud}{\ud t} \mathrm{tr}[\rho^{\alpha}(t) + \rho^{\beta}(t)] = \int_{\mathbb{R}^d} \ud p\, \mathrm{tr}^{\alpha\beta}[\mathcal{C}[W](p,t)] = 0
\end{equation}
for $\alpha\beta = \ptype{ab}$, $\ptype{cd}$, $\ptype{ad}$ or $\ptype{cb}$. The trace $\mathrm{tr}^{\alpha\beta}$ is understood to act on the blocks $\alpha$ and $\beta$ only, i.e.,
\begin{equation}
\mathrm{tr}^{\alpha\beta}[W(p,t)] = \mathrm{tr}[W^\alpha(p,t)] + \mathrm{tr}[W^\beta(p,t)].
\end{equation}
Relation~\eqref{eq:DensitySubsetCons} holds since the integrand of the dissipative $\mathcal{C}_{\mathrm{diss}}$ vanishes after appropriate interchange of $1$, $2$, $3$, $4$: note that
\begin{equation}
\sum_{\sigma \in \{\uparrow, \downarrow\}} \{ W^{\alpha}_i, \ket{\sigma}\bra{\sigma} \} = 2 \, W^{\alpha}_i,
\end{equation}
such that for $\alpha\beta = \ptype{ab}$, say, the traces of the $\ptype{a}$-component in Eq.~\eqref{eq:AATa_tensor} and $\ptype{b}$-component in Eq.~\eqref{eq:AATb_tensor} (with $(1,3) \leftrightarrow (2,4)$) cancel out. The conservative collision operator $\mathcal{C}_{\mathrm{cons}}$ inserted into \eqref{eq:DensitySubsetCons} vanishes immediately since $\mathcal{C}_{\mathrm{cons}}$ is a commutator.

Note that taking the trace is indeed required in Eq.~\eqref{eq:DensityCons}, i.e., the individual spin components are not conserved in general.

\subsection{Momentum conservation}
\label{sec:MomentumConversation}

Momentum conservation
\begin{equation}
\label{eq:MomentumCons}
\frac{\ud}{\ud t} \int_{\mathbb{R}^d} \ud p\, p\, \mathrm{tr}\big[W(p,t)\big] = \int_{\mathbb{R}^d} \ud p\, p\, \mathrm{tr}\big[\mathcal{C}[W(p,t)]\big] = 0
\end{equation}
follows from the factor $\underline{p}\,\delta(\underline{p})$ in the integrand after appropriate interchanges $1 \leftrightarrow 3$, $2 \leftrightarrow 4$ and $(1,3) \leftrightarrow (2,4)$. Isotropic states always have zero average momentum.

\subsection{Energy conservation}
\label{sec:EnergyConversation}

Energy conservation is represented by the equation
\begin{equation}
\begin{split}
\label{eq:EnergyCons}
&\frac{\ud}{\ud t} \int_{\mathbb{R}^d} \ud p\, \mathrm{tr}\big[\omega(p) \cdot W(p,t)\big] \\
&= \int_{\mathbb{R}^d} \ud p\, \mathrm{tr}\big[\omega(p) \cdot \mathcal{C}[W(p,t)]\big] = 0
\end{split}
\end{equation}
with the dispersion matrix $\omega(p)$ defined in Eq.~\eqref{eq:omegaDef}. The term inside the trace is a $8 \times 8$ matrix. Similar to the momentum conservation, Eq.~\eqref{eq:EnergyCons} follows from the factor $(\omega^\ptype{a}_1 - \omega^\ptype{b}_2 + \omega^\ptype{c}_3 - \omega^\ptype{d}_4)\, \delta(\omega^\ptype{a}_1 - \omega^\ptype{b}_2 + \omega^\ptype{c}_3 - \omega^\ptype{d}_4)$ in the integrand after appropriate interchanges $1 \leftrightarrow 3$, $2 \leftrightarrow 4$ and $(1,3) \leftrightarrow (2,4)$.

\subsection{Additional conservation laws depending on the interaction matrices}
\label{sec:AdditionalConservation}

Taking all conservation laws into account is necessary for computing the asymptotic (thermal) equilibrium state (see Sec.~\ref{sec:Stationary} below), and there are additional conservation laws depending on the $V^{\alpha \beta}$ matrices. Since the collision operator can be expressed in terms of the $\mathcal{V}$ matrix in Eq.~\eqref{eq:Vop}, it suffices to discuss the structure and zero pattern of the entries of $\mathcal{V}$, which is to be understood modulo unitary rotations of the form \eqref{eq:unitary_invariance}. Whenever such rotations lead to a particular pattern as discussed in the following, the respective conservation law holds in this basis.

We will only consider $V^{\alpha \beta}$ matrices with full rank 2, to exclude degenerate cases like $V^{\alpha \beta} = 0$ (as a matrix).

\paragraph{General diagonal $V^{\alpha \beta}$.}
The $\mathcal{V}$ matrix represented in the standard basis $(\ket{\uparrow\uparrow}$, $\ket{\uparrow\downarrow}$, $\ket{\downarrow\uparrow}$, $\ket{\downarrow\downarrow})$ has the structure
\begin{equation}
\label{eq:diagVop_zeros}
\mathcal{V} = \begin{pmatrix}
* & 0 & 0 & 0 \\
0 & * & * & 0 \\
0 & * & * & 0 \\
0 & 0 & 0 & *
\end{pmatrix},
\end{equation}
where each star represents an arbitrary number. In this case, the diagonal entries of the total spin remain constant under the time evolution of the Boltzmann equation,
\begin{equation}
\label{eq:diag_spin}
\frac{\ud}{\ud t} \rho_{\sigma\sigma}(t) = \frac{\ud}{\ud t} \sum_{\alpha \in \{\ptype{a}, \ptype{b}, \ptype{c}, \ptype{d}\}} \int_{\mathbb{R}^d} \ud p\, W^{\alpha}_{\sigma\sigma}(p,t) = 0.
\end{equation}
To prove this assertion, consider the $\uparrow\uparrow$ entry (the proof for the $\downarrow\downarrow$ entry proceeds analogously). Expanding the representation \eqref{eq:AATa_tensor} gives
\begin{equation}
\label{eq:AAT_proj_up}
\begin{split}
&\sum_{\alpha \in \{\ptype{a}, \ptype{b}, \ptype{c}, \ptype{d}\}} \bra{\uparrow} \big( \mathcal{A}_{\mathrm{quad}}[W]_{1234} + \mathcal{A}_{\mathrm{tr}}[W]_{1234} \big)^{\alpha} \ket{\uparrow} \\
&= \sum_{\boldsymbol{\sigma}, \boldsymbol{\tau}}\, D(\boldsymbol{\sigma}, \boldsymbol{\tau}) \times\\
&\quad \big( \ \bra{\sigma_1 \sigma_3} \big( \tilde{W}^\ptype{a}_1 \otimes \tilde{W}^\ptype{c}_3 \big) \ket{\tau_1 \tau_3} \bra{\sigma_2 \sigma_4} \big(W^\ptype{b}_2 \otimes W^\ptype{d}_4\big) \ket{\tau_2 \tau_4} \\
&\hspace{7pt} - \bra{\sigma_1 \sigma_3} \big( W^\ptype{a}_1 \otimes W^\ptype{c}_3 \big) \ket{\tau_1 \tau_3} \bra{\sigma_2 \sigma_4} \big(\tilde{W}^\ptype{b}_2 \otimes \tilde{W}^\ptype{d}_4\big) \ket{\tau_2 \tau_4} \big) \\
&\qquad \times \bra{\tau_1 \tau_3} \mathcal{V} \ket{\sigma_2 \sigma_4} \bra{\tau_2 \tau_4} \mathcal{V}^* \ket{\sigma_1 \sigma_3}
\end{split}
\end{equation}
with
\begin{equation}
\begin{split}
D(\boldsymbol{\sigma}, \boldsymbol{\tau})
&= \big(\delta_{\sigma_1,\uparrow} + \delta_{\tau_1,\uparrow} - \delta_{\sigma_2,\uparrow} - \delta_{\tau_2,\uparrow} \\
&\hspace{5pt} + \delta_{\sigma_3,\uparrow} + \delta_{\tau_3,\uparrow} - \delta_{\sigma_4,\uparrow} - \delta_{\tau_4,\uparrow} \big)
\end{split}
\end{equation}
and the notation $\boldsymbol{\sigma} = (\sigma_1, \sigma_2, \sigma_3, \sigma_4)$, $\boldsymbol{\tau} = (\tau_1, \tau_2, \tau_3, \tau_4)$. Direct inspection shows that $D(\boldsymbol{\sigma}, \boldsymbol{\tau}) = 0$ or $\bra{\tau_1 \tau_3} \mathcal{V} \ket{\sigma_2 \sigma_4} \bra{\tau_2 \tau_4} \mathcal{V}^* \ket{\sigma_1 \sigma_3} = 0$ for all spin combinations, given the zero pattern in Eq.~\eqref{eq:diagVop_zeros}.

There are $5 + d$ independently conserved quantities: the two diagonal entries in Eq.~\eqref{eq:diag_spin}, the densities of $\ptype{a} + \ptype{b}$ and $\ptype{a} + \ptype{d}$ according to Eq.~\eqref{eq:DensitySubsetCons}, the momentum and the total energy. The other conserved quantities are redundant; for example, the density of $\ptype{c} + \ptype{d}$ can be obtained from the sum of the diagonal entries in Eq.~\eqref{eq:diag_spin} minus the density of $\ptype{a} + \ptype{b}$.

\paragraph{All $V^{\alpha \beta}$ proportional to the identity matrix.}
This is a special case of (a), relevant for the $\beta$ decay discussed below, and $\mathcal{V}$ is of the form $\mathcal{V} = c^{\scriptscriptstyle{=}}\,\mathbbm{1}_{4 \times 4} + c^{\mathsf{x}}\,T$ with two constants $c^{\scriptscriptstyle{=}}$ and $c^{\mathsf{x}}$. In this case $\mathcal{V}$ is invariant under a simultaneous unitary rotation of the Wigner matrices as in Eq.~\eqref{eq:unitary_invariance} with $U^{\ptype{a}} = U^{\ptype{b}} = U^{\ptype{c}} = U^{\ptype{d}} = U \in \mathrm{SU}(2)$, i.e., $(U \otimes U)\,\mathcal{V}\,(U \otimes U)^* = \mathcal{V}$. Such a simultaneous rotation sends $\rho(t) \to U \rho(t) U^*$, and together with Eq.~\eqref{eq:diag_spin}, it follows that the total spin density matrix remains constant in time,
\begin{equation}
\label{eq:rho_t_constant}
\frac{\ud}{\ud t} \rho(t) = 0.
\end{equation}
Alternatively, one could prove this assertion starting directly from Eq.~\eqref{eq:AATa_tensor}, together with the identities $\mathrm{tr}[A \otimes B] = \mathrm{tr}[A]\,\mathrm{tr}[B]$ and $\mathrm{tr}[\{A,C\}\cdot B] = \mathrm{tr}[A\cdot\{B,C\}]$, which are valid for any matrices $A$, $B$ and $C$.

\paragraph{Zero outer frame in $\mathcal{V}$ matrix.}
We investigate the zero pattern
\begin{equation}
\label{eq:Vop_zeroframe}
\mathcal{V} = \begin{pmatrix}
0 & 0 & 0 & 0 \\
0 & * & * & 0 \\
0 & * & * & 0 \\
0 & 0 & 0 & 0
\end{pmatrix},
\end{equation}
represented in the standard basis $(\ket{\uparrow\uparrow}$, $\ket{\uparrow\downarrow}$, $\ket{\downarrow\uparrow}$, $\ket{\downarrow\downarrow})$ as above. This pattern can emerge from non-diagonal $V^{\alpha \beta}$ interaction matrices with full rank, too.
%
Besides the conservation of the diagonal entries in Eq.~\eqref{eq:diag_spin}, the projection onto the Pauli matrix $\sigma_z = \bigl(\begin{smallmatrix} 1 & 0 \\ 0 & -1\end{smallmatrix}\bigr)$ for types $\ptype{a} + \ptype{c}$ and $\ptype{b} + \ptype{d}$ is also conserved, i.e.,
\begin{equation}
\label{eq:sigma_z_spin}
\frac{\ud}{\ud t} \sum_{\alpha} \mathrm{tr}[\sigma_z\, \rho^{\alpha}(t)] = \frac{\ud}{\ud t} \sum_{\alpha} \int_{\mathbb{R}^d} \ud p\, \mathrm{tr}[\sigma_z\, W^{\alpha}(p,t)] = 0
\end{equation}
with summation over $\alpha \in \{\ptype{a}, \ptype{c}\}$ or $\alpha \in \{\ptype{b}, \ptype{d}\}$. To prove this statement, first note that
\begin{equation}
\{ W^{\alpha}_i, \sigma_z \} = 2 \begin{pmatrix} \bra{\uparrow} W^{\alpha}_i \ket{\uparrow} & 0 \\ 0 & -\bra{\downarrow} W^{\alpha}_i \ket{\downarrow} \end{pmatrix}.
\end{equation}
Then we proceed as for diagonal $V^{\alpha \beta}$ above, except that  $D(\boldsymbol{\sigma},\boldsymbol{\tau})$ in Eq.~\eqref{eq:AAT_proj_up} is replaced by
\begin{equation}
D'(\boldsymbol{\sigma}, \boldsymbol{\tau})
= 2 \big( \bra{\sigma_1} \sigma_z \ket{\tau_1} + \bra{\sigma_3} \sigma_z \ket{\tau_3} \big)
\end{equation}
for $\alpha \in \{\ptype{a}, \ptype{c}\}$. As before, $D'(\boldsymbol{\sigma}, \boldsymbol{\tau}) = 0$ or $\bra{\tau_1 \tau_3} \mathcal{V} \ket{\sigma_2 \sigma_4} \bra{\tau_2 \tau_4} \mathcal{V}^* \ket{\sigma_1 \sigma_3} = 0$ for all spin combinations, given the pattern in Eq.~\eqref{eq:Vop_zeroframe}.

In summary, there are $6 + d$ independently conserved quantities: the $5 + d$ quantities from case (a) with diagonal $V^{\alpha \beta}$, and the projection onto $\sigma_z$ in Eq.~\eqref{eq:sigma_z_spin} with summation over $\alpha \in \{ \ptype{a}, \ptype{c} \}$. Summation over $\alpha \in \{ \ptype{b}, \ptype{d} \}$ is redundant due to Eq.~\eqref{eq:diag_spin}.

\begin{table*}[!ht]
\centering
\begin{tabular}{>{\centering\arraybackslash}m{4cm} | >{\centering\arraybackslash}m{2.4cm} >{\centering\arraybackslash}m{2.3cm} >{\centering\arraybackslash}m{2.3cm} >{\centering\arraybackslash}m{1.3cm} >{\centering\arraybackslash}m{1.3cm} >{\centering\arraybackslash}m{2.8cm} @{}m{0pt}@{}}
structure of $\mathcal{V}$ & \multicolumn{5}{c}{conserved quantities} \\[2ex]
\hline
general $\mathcal{V}$ & momentum \eqref{eq:MomentumCons} and energy \eqref{eq:EnergyCons} & $\mathrm{tr}[\rho^{\ptype{a}}(t) + \rho^{\ptype{b}}(t)]$ & $\mathrm{tr}[\rho^{\ptype{a}}(t) + \rho^{\ptype{d}}(t)]$ & $\mathrm{tr}[\rho(t)]$ & & & \\[4ex]
\hline
$\mathcal{V}$ in Eq.~\eqref{eq:diagVop_zeros} (general diagonal $V^{\alpha \beta}$) & $\shortparallel$ & $\shortparallel$ & $\shortparallel$ & $\rho_{\uparrow\uparrow}(t)$ & $\rho_{\downarrow\downarrow}(t)$ & & \\[4ex]
\hline
$\mathcal{V} = c^{\scriptscriptstyle{=}}\,\mathbbm{1}_{4 \times 4} + c^{\mathsf{x}}\,T$ ($V^{\alpha \beta}$ proportional to identity) & $\shortparallel$ & $\shortparallel$ & $\shortparallel$ & $\rho(t)$ & & & \\[4ex]
\hline
zero outer frame in $\mathcal{V}$ matrix (Eq.~\eqref{eq:Vop_zeroframe}) & $\shortparallel$ & $\shortparallel$ & $\shortparallel$ & $\rho_{\uparrow\uparrow}(t)$ & $\rho_{\downarrow\downarrow}(t)$ & $\mathrm{tr}[\sigma_z \big(\rho^{\ptype{a}}(t) + \rho^{\ptype{c}}(t)\big) ]$ & \\[4ex]
\hline
\end{tabular}
\caption{Independently conserved quantities, depending on the special structure of $\mathcal{V}$}
\label{tab:ConservedQuantities}
\end{table*}
The independently conserved quantities are summarized in table~\ref{tab:ConservedQuantities}.

\subsection{H-theorem}
\label{sec:Htheorem}

In the following, we prove the H-theorem which states that the entropy is monotonically increasing. We represent each Wigner function by its spectral decomposition
\begin{equation}
\label{eq:spectral_decomp}
W^{\alpha}(p) = \sum_{\sigma \in \{\uparrow, \downarrow\}} \lambda^{\alpha}_\sigma(p) \, P^{\alpha}_\sigma(p)
\end{equation}
for $\alpha \in \{\ptype{a}, \ptype{b}, \ptype{c}, \ptype{d}\}$, where $0 \le \lambda^{\alpha}_\sigma(p) \le 1$ are the eigenvalues and $P^{\alpha}_\sigma(p) = \ket{\alpha; p, \sigma} \bra{\alpha; p,\sigma}$ an orthogonal eigenprojector.

The entropy production is given by
\begin{equation}
\label{eq:sigmaW}
\sigma[W] = \frac{\ud}{\ud t} S[W] = -\int_{\mathbb{R}^3} \ud p_1 \, \mathrm{tr}[ (\log W_1 - \log \tilde{W}_1) \, \mathcal{C}[W]_1 ].
\end{equation}
%
%
In the following, we will use the shorthand notation $\lambda_j = \lambda^{(j)}_{\sigma_j}(p_j)$, $P_j = P^{(j)}_{\sigma_j}(p_j) = \lvert j\rangle\langle j\rvert$, and $\sum_{\boldsymbol{\sigma}} = \sum_{\sigma_1,\sigma_2,\sigma_3,\sigma_4}$. For example, $\lambda_2 = \lambda^{\ptype{b}}_{\sigma_2}(p_2)$. Inserting the spectral decomposition~\eqref{eq:spectral_decomp} and the integrand representation~\eqref{eq:AATa_tensor} of the dissipative collision operator into Eq.~\eqref{eq:sigmaW}, the contribution of the $(\ptype{a},\ptype{a})$-component (first $2 \times 2$ block) to the entropy production reads
\begin{equation}
\label{eq:sigmaW_aa}
\begin{split}
&\sigma[W]^{\ptype{a}} = - \pi \int \ud p_{1234} \, \delta(\underline{p}) \, \delta\!\left(\omega^{\ptype{a}}_1 - \omega^{\ptype{b}}_2 + \omega^{\ptype{c}}_3 - \omega^{\ptype{d}}_4\right) \\
&\ \times \mathrm{tr}\Big[ \big( \{ \tilde{W}^\ptype{a}_1, \log W_1 - \log \tilde{W}_1 \} \otimes \tilde{W}^\ptype{c}_3 \big) \, \mathcal{V} \, \big(W^\ptype{b}_2 \otimes W^\ptype{d}_4\big) \, \mathcal{V}^* \\
&\hspace{16pt} - \big( \{ W^\ptype{a}_1, \log W_1 - \log \tilde{W}_1 \} \otimes W^\ptype{c}_3 \big) \, \mathcal{V} \, \big(\tilde{W}^\ptype{b}_2 \otimes \tilde{W}^\ptype{d}_4\big) \, \mathcal{V}^*
\Big] \\
&= 2\pi \int \ud p_{1234} \, \delta(\underline{p}) \, \delta\!\left(\omega^{\ptype{a}}_1 - \omega^{\ptype{b}}_2 + \omega^{\ptype{c}}_3 - \omega^{\ptype{d}}_4\right) \\
&\quad \times \sum_{\boldsymbol{\sigma}} \left(\log \lambda_1 - \log\tilde{\lambda}_1 \right) \left( \lambda_1 \tilde{\lambda}_2 \lambda_3 \tilde{\lambda}_4 - \tilde{\lambda}_1 \lambda_2 \tilde{\lambda}_3 \lambda_4 \right) \\
&\quad \times \bra{1 3} \mathcal{V} \ket{2 4} \bra{2 4} \mathcal{V}^* \ket{1 3}.
\end{split}
\end{equation}
The contribution of the $(\ptype{b},\ptype{b})$-component to the entropy production coincides with Eq.~\eqref{eq:sigmaW_aa} after permuting $\ptype{a} \leftrightarrow \ptype{b}$, $\ptype{c} \leftrightarrow \ptype{d}$. Together with relabeling the integration variables $1 \leftrightarrow 2$ and $3 \leftrightarrow 4$, the contribution of the $(\ptype{b},\ptype{b})$-component has exactly the same form as \eqref{eq:sigmaW_aa} upon replacing
\begin{equation}
\left(\log \lambda_1 - \log\tilde{\lambda}_1 \right) \to - \left(\log \lambda_2 - \log\tilde{\lambda}_2 \right).
\end{equation}
Similar reasoning holds for the contributions from the $(\ptype{c},\ptype{c})$ and $(\ptype{d},\ptype{d})$ components. In summary, the entropy production equals
\begin{equation}
\label{eq:sigmaWge0}
\begin{split}
&\sigma[W] = 2\pi \int \ud p_{1234} \, \delta(\underline{p}) \, \delta\!\left(\omega^{\ptype{a}}_1 - \omega^{\ptype{b}}_2 + \omega^{\ptype{c}}_3 - \omega^{\ptype{d}}_4\right) \\
&\quad \times \sum_{\boldsymbol{\sigma}} \log\!\left( \frac{\lambda_1 \tilde{\lambda}_2 \lambda_3 \tilde{\lambda}_4}{\tilde{\lambda}_1 \lambda_2 \tilde{\lambda}_3 \lambda_4} \right) \left( \lambda_1 \tilde{\lambda}_2 \lambda_3 \tilde{\lambda}_4 - \tilde{\lambda}_1 \lambda_2 \tilde{\lambda}_3 \lambda_4 \right) \\
&\quad \times \abs{\bra{1 3} \mathcal{V} \ket{2 4} }^2 \ge 0
\end{split}
\end{equation}
since $(x - y) \log(x/y) \ge 0$.

\section{Stationary states}
\label{sec:Stationary}

\setcounter{paragraph}{0}

All stationary states have to satisfy $\sigma[W] = 0$, i.e., the entropy production must be zero. To elucidate the set of Wigner functions which adhere to this condition, we define (in the context of the proof of the H-theorem)
\begin{equation}
\label{eq:Fdef}
F(\boldsymbol{p},\boldsymbol{\sigma}) = \log\!\left( \frac{\lambda_1 \tilde{\lambda}_2 \lambda_3 \tilde{\lambda}_4}{\tilde{\lambda}_1 \lambda_2 \tilde{\lambda}_3 \lambda_4} \right) \left( \lambda_1 \tilde{\lambda}_2 \lambda_3 \tilde{\lambda}_4 - \tilde{\lambda}_1 \lambda_2 \tilde{\lambda}_3 \lambda_4 \right) \ge 0
\end{equation}
and $\mathcal{V}(\boldsymbol{p},\boldsymbol{\sigma}) = \bra{1 3} \mathcal{V} \ket{2 4}$, where $\mathcal{V}$ is the $4 \times 4$ matrix in Eq.~\eqref{eq:Vop} and we have used the notation $\lvert j\rangle\langle j\rvert = P^{(j)}_{\sigma_j}(p_j)$ from above. It must hold that $F(\boldsymbol{p},\boldsymbol{\sigma}) = 0$ or $\mathcal{V}(\boldsymbol{p},\boldsymbol{\sigma}) = 0$ (or both) for each configuration of the $\boldsymbol{\sigma}$ variables, according to Eq.~\eqref{eq:sigmaWge0}. Defining the collision invariants as
\begin{equation}
\label{eq:phi_definition}
\Phi_\sigma^{\alpha}(p) = \log\left( \frac{\lambda_\sigma^{\alpha}(p)}{\tilde{\lambda}_\sigma^{\alpha}(p)} \right),
\end{equation}
then $F(\boldsymbol{p},\boldsymbol{\sigma}) = 0$ is equivalent to
\begin{equation}
\label{eq:phi_condition}
\Phi_{\sigma_1}^\ptype{a}(p_1) - \Phi^\ptype{b}_{\sigma_2}(p_2) + \Phi^\ptype{c}_{\sigma_3}(p_3) - \Phi^\ptype{d}_{\sigma_4}(p_4) = 0.
\end{equation}
Based on general arguments \cite{CollisionalInvariants2006}, one expects that the Wigner functions will equilibrate as $t \to \infty$, i.e., converge to thermal equilibrium (Fermi-Dirac) distributions
\begin{equation}
\label{eq:FermiDirac}
\begin{split}
W^{\alpha}_{\mathrm{eq}}(p) &= \sum_{\sigma \in \{\uparrow, \downarrow\}} \lambda^{\alpha}_{\mathrm{eq},\sigma}(p) \ket{\alpha;\sigma} \bra{\alpha;\sigma} \quad \text{with} \\
\lambda^{\alpha}_{\mathrm{eq},\sigma}(p) &= \left(\mathrm{e}^{\beta\, \left(\omega^{\alpha}(p) - \mu_\sigma^{\alpha}\right)} + 1\right)^{-1}.
\end{split}
\end{equation}
Here we have assumed that the orthonormal eigenbasis $\ket{\alpha;\sigma}$ is independent of $p$ (thus $\mathcal{V}(\boldsymbol{p},\boldsymbol{\sigma}) = \mathcal{V}(\boldsymbol{\sigma})$), that the average momentum is zero, and that all particle types share the same inverse temperature $\beta$. We exclude degenerate cases like $V^{\alpha\beta} = 0$ as a matrix. Inserting the Fermi-Dirac eigenvalues in \eqref{eq:FermiDirac} into \eqref{eq:phi_condition} and using the energy conservation translates to the \emph{linear} equation
\begin{equation}
\label{eq:mu_condition}
F(\boldsymbol{p},\boldsymbol{\sigma}) = 0 \quad \Leftrightarrow \quad \mu_{\sigma_1}^{\ptype{a}} - \mu_{\sigma_2}^{\ptype{b}} + \mu_{\sigma_3}^{\ptype{c}} - \mu_{\sigma_4}^{\ptype{d}} = 0.
\end{equation}

The remaining task is to determine the chemical potentials $\mu_\sigma^{\alpha}$, inverse temperature $\beta$ and the basis $\ket{\alpha;\sigma}$ in accordance with the conservation laws, which themselves depend on $\mathcal{V}$.

For the following, it is convenient to represent the right side of Eq.~\eqref{eq:mu_condition} for each $\boldsymbol{\sigma}$ combination as $4 \times 4$ matrix (denoted by $\mathcal{F}$) with entries
\begin{equation}
\bra{\sigma_1 \sigma_3} \mathcal{F} \ket{\sigma_2 \sigma_4} = \mu_{\sigma_1}^{\ptype{a}} - \mu_{\sigma_2}^{\ptype{b}} + \mu_{\sigma_3}^{\ptype{c}} - \mu_{\sigma_4}^{\ptype{d}}
\end{equation}
with respect to the standard basis $( \ket{\uparrow\uparrow}, \ket{\uparrow\downarrow}, \ket{\downarrow\uparrow}, \ket{\downarrow\downarrow} )$. This representation is analogous to the $\mathcal{V}$ matrix.

After changing basis according to Eq.~\eqref{eq:unitary_invariance}, $\mathcal{V}$ may be represented in the eigenbasis $\ket{\alpha;\sigma}$, and we can without loss of generality assume that $\ket{\alpha;\sigma}$ is the standard basis.

In what follows, we discuss a (non-exhaustive) list of special cases (as for the additional conservation laws in Sec.~\ref{sec:AdditionalConservation}).

\paragraph{General $\mathcal{V}$.}
We assume that $\mathcal{V}$ exhibits none of the zero patterns below, even after unitary rotations of the form \eqref{eq:unitary_invariance}. Explicit enumeration using a computer algebra system shows the following: whenever the condition \eqref{eq:mu_condition} holds for at least 9 (pairwise different) configurations of the $\boldsymbol{\sigma}$ variables, then all chemical potentials are necessarily independent of spin, 
\begin{equation}
\label{eq:mu_basic_sol}
\mu_{\sigma}^{\alpha} = \nu^{\alpha} \quad \text{with} \quad \nu^{\ptype{a}} - \nu^{\ptype{b}} + \nu^{\ptype{c}} - \nu^{\ptype{d}} = 0.
\end{equation}
In this case $F(\boldsymbol{p},\boldsymbol{\sigma}) = 0$ always. According to the first row in table~\ref{tab:ConservedQuantities}, there are $4$ independently conserved quantities (for zero average momentum), and correspondingly $4$ parameters to describe the equilibrium state, namely $\beta$, $\nu^{\ptype{a}}$, $\nu^{\ptype{b}}$ and $\nu^{\ptype{c}}$ ($\nu^{\ptype{d}}$ is fixed by Eq.~\eqref{eq:mu_basic_sol}). Note that the choice of the basis $\ket{\alpha;\sigma}$ is arbitrary in the present case due to independence of spin.

\paragraph{$\mathcal{V}$ with zero structure in Eq.~\eqref{eq:diagVop_zeros}.}
This case is equivalent to general diagonal $V^{\alpha \beta}$ matrices. Since $F(\boldsymbol{p},\boldsymbol{\sigma}) = 0$ must hold whenever $\mathcal{V}(\boldsymbol{\sigma}) \neq 0$, the required complementary zero pattern for $\mathcal{F}$ reads
\begin{equation}
\mathcal{F} =
\begin{pmatrix}
0 & * & * & * \\
* & 0 & 0 & * \\
* & 0 & 0 & * \\
* & * & * & 0 \\
\end{pmatrix}.
\end{equation}
Solving the linear equations \eqref{eq:mu_condition} corresponding to the zero entries of this matrix leads to the solution
\begin{equation*}
\mu_{\sigma}^{\alpha} = \nu^{\alpha} + c \bra{\sigma} \sigma_z \ket{\sigma}
\end{equation*}
with a fixed $c \in \mathbb{R}$ and $\nu^{\ptype{a}} - \nu^{\ptype{b}} + \nu^{\ptype{c}} - \nu^{\ptype{d}} = 0$. There are $5$ independent parameters (in accordance with the $5$ conservation laws in the second row of table~\ref{tab:ConservedQuantities}): the values of $\beta$, $\nu^{\ptype{a}}$, $\nu^{\ptype{b}}$, $\nu^{\ptype{c}}$ and $c$.

\paragraph{$\mathcal{V} = c^{\scriptscriptstyle{=}}\,\mathbbm{1}_{4 \times 4} + c^{\mathsf{x}}\,T$.}
This structure results from all $V^{\alpha \beta}$ matrices proportional to the identity matrix, summarized in the third row of table~\ref{tab:ConservedQuantities}. Since $\rho(t)$ remains constant in time, we can diagonalize $\rho(t)$ by a global, constant unitary rotation $U \in \mathrm{SU}(2)$. Thus, without loss of generality one can assume that $\rho(t)$ is diagonal. From here the argumentation proceeds as in the previous case with general diagonal $V^{\alpha \beta}$.

\paragraph{Zero outer frame in $\mathcal{V}$ matrix, Eq.~\eqref{eq:Vop_zeroframe}.}
The complementary zero pattern for $\mathcal{F}$ is
\begin{equation}
\mathcal{F} = \begin{pmatrix}
* & * & * & * \\
* & 0 & 0 & * \\
* & 0 & 0 & * \\
* & * & * & *
\end{pmatrix}.
\end{equation}
Solving the corresponding system of linear equations according to \eqref{eq:mu_condition} leads to 
\begin{equation}
\label{eq:mu_zero4}
\mu_{\sigma}^{\alpha} = \nu^{\alpha} + c^{\alpha} \bra{\sigma} \sigma_z \ket{\sigma}
\end{equation}
with $c^{\ptype{a}} = c^{\ptype{c}}$, $c^{\ptype{b}} = c^{\ptype{d}}$ and $\nu^{\ptype{a}} - \nu^{\ptype{b}} + \nu^{\ptype{c}} - \nu^{\ptype{d}} = 0$. The number of independent parameters ($\beta$, $\nu^{\ptype{a}}$, $\nu^{\ptype{b}}$, $\nu^{\ptype{c}}$, $c^{\ptype{a}}$ and $c^{\ptype{b}}$) for zero average momentum matches the number of conserved quantities, see last row in table~\ref{tab:ConservedQuantities}.

In practice, we fit $\beta$ and the additional parameters numerically such that the conserved quantities obtained from the corresponding Fermi-Dirac state match the ones of the initial state. We conjecture that the map from the conserved quantities to the parameters is one to one.

\begin{figure*}[!ht]
\centering
\subfloat[$W^{\ptype{a}}(p,0)$ with $m^{\ptype{a}} = 1$]{
\includegraphics[width=0.5\columnwidth]{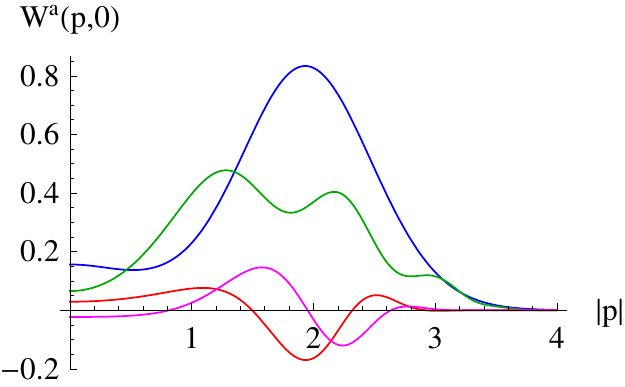}
\label{fig:Wa0matrix}} 
\subfloat[$W^{\ptype{b}}(p,0)$ with $m^{\ptype{b}} = \frac{4}{5}$]{
\includegraphics[width=0.5\columnwidth]{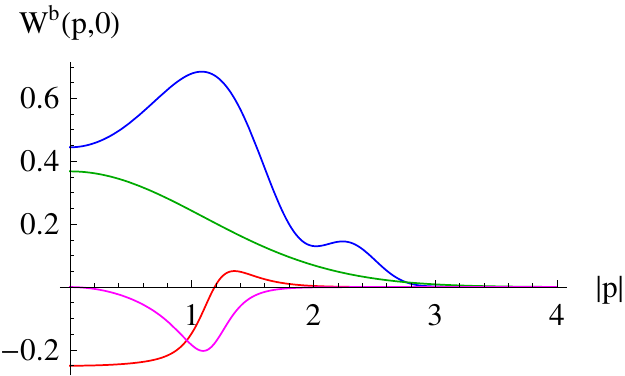}
\label{fig:Wb0matrix}} 
\subfloat[$W^{\ptype{c}}(p,0)$ with $m^{\ptype{c}} = \frac{1}{5}$]{
\includegraphics[width=0.5\columnwidth]{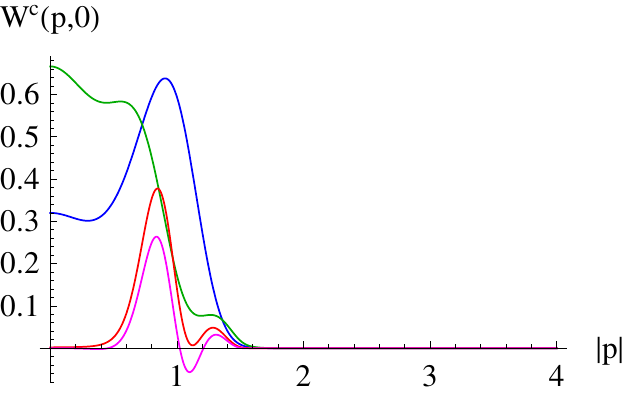}
\label{fig:Wc0matrix}} 
\subfloat[$W^{\ptype{d}}(p,0)$ with $m^{\ptype{d}} = \frac{1}{2}$]{
\includegraphics[width=0.5\columnwidth]{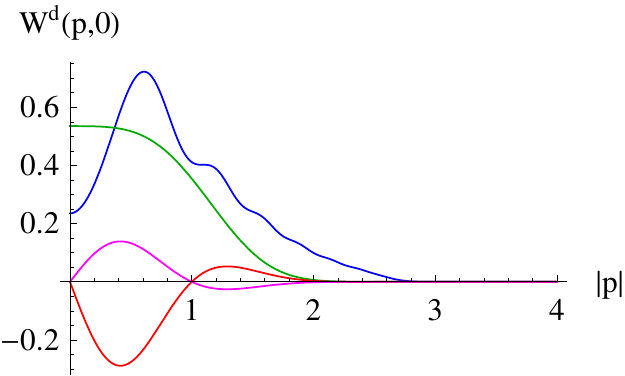}
\label{fig:Wd0matrix}} \\
\subfloat[eigenvalues of $W^{\ptype{a}}(p,0)$]{
\includegraphics[width=0.5\columnwidth]{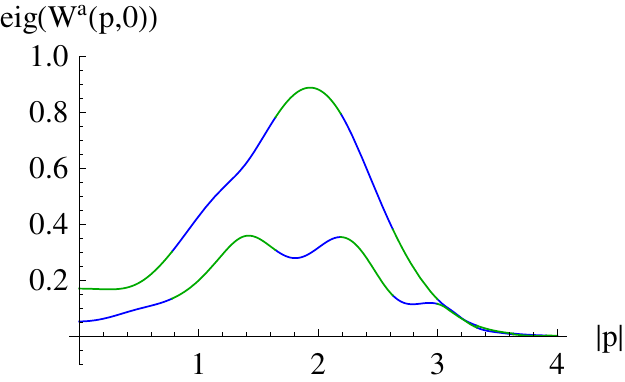}
\label{fig:Wa0eig}}
\subfloat[eigenvalues of $W^{\ptype{b}}(p,0)$]{
\includegraphics[width=0.5\columnwidth]{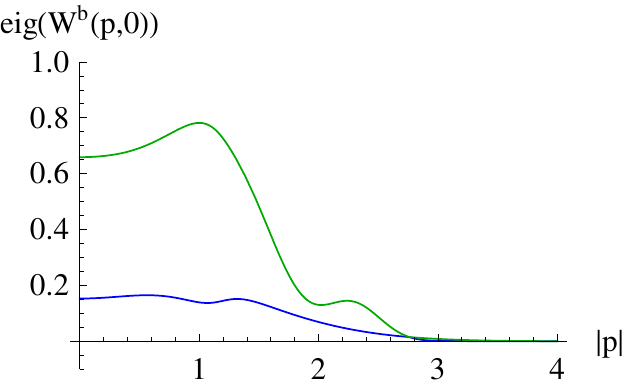}
\label{fig:Wb0eig}}
\subfloat[eigenvalues of $W^{\ptype{c}}(p,0)$]{
\includegraphics[width=0.5\columnwidth]{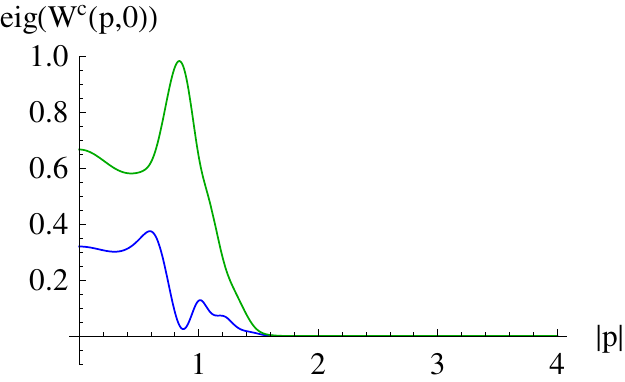}
\label{fig:Wc0eig}}
\subfloat[eigenvalues of $W^{\ptype{d}}(p,0)$]{
\includegraphics[width=0.5\columnwidth]{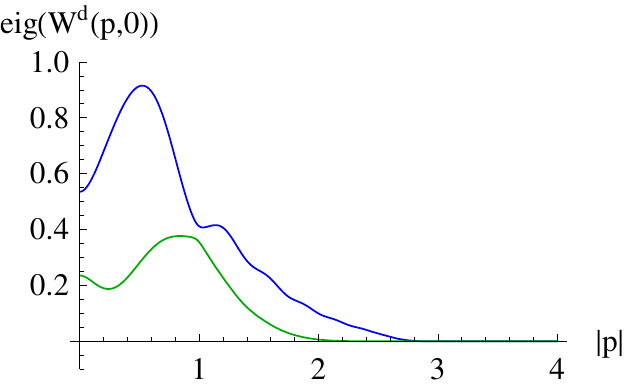}
\label{fig:Wd0eig}}
\caption{(Color online) The initial state $W(p,0)$ used for the simulations. (a) -- (d) Matrix entries: the blue and green (upper) curves show the real diagonal entries, and the red and magenta curves the real and imaginary parts of the off-diagonal $\ket{\uparrow}\bra{\downarrow}$ entry, respectively. (e) -- (h) Corresponding eigenvalues of $W^{\alpha}(p,0)$.}
\label{fig:W0}
\end{figure*}

\section{Numerical Procedure}
\label{sec:Numerics}

Concerning the numeric integration for the dissipative collision operator, our goal is to solve the following $p_1$-dependent integral numerically:
\begin{multline}
\label{eq:coll_int}
\mathcal{C}_{\mathrm{diss}}[W]^{\alpha}_1 = \pi \int_{\mathbb{R}^9} \ud p_{234} \, \delta^3(\underline{p}) \, \delta(\omega^{\alpha\beta\gamma\delta}) \\
\times \left( \mathcal{A}_{\mathrm{quad}}[W]_{1234} + \mathcal{A}_{\mathrm{tr}}[W]_{1234} \right)^{\alpha},
\end{multline}
where we have used the notation $\underline{p} = p_1 - p_2 + p_3 - p_4$ and $\omega^{\alpha\beta\gamma\delta} = \omega^\alpha_1 - \omega^\beta_2 + \omega^\gamma_3 - \omega^\delta_4$.

We follow the derivation~\cite[appendix~A]{SemikozTkachev1997} to resolve the $\delta$-functions in the collision integral~\eqref{eq:coll_int} as far as possible and to integrate out the angular parts. Expressed in terms of the energies $\varepsilon_i = \abs{p_i}^2 / (2 m^{i})$ with $m^1 = m^{\alpha}$, $m^2 = m^{\beta}$ etc., and using the relation
\begin{equation}
\ud^3 p_i = \ud \Omega\, \abs{p_i} \, m^i\, \ud\varepsilon_i,
\end{equation}
one arrives at the following two-dimensional integral:
\begin{multline}
\label{eq:coll_int_en}
\mathcal{C}_{\mathrm{diss}}[W]^{\alpha}_1 = (2\pi)^3 \, m^\beta m^\gamma m^\delta \\
\times \int_{\mathcal{D}(\varepsilon_1)} \ud \varepsilon_2 \, \ud \varepsilon_4 \, \frac{\min(\abs{p_1},\abs{p_2},\abs{p_3},\abs{p_4})}{\abs{p_1}} \\
\times \left( \mathcal{A}_{\mathrm{quad}}[W]_{1234} + \mathcal{A}_{\mathrm{tr}}[W]_{1234} \right)^{\alpha}
\end{multline}
with the integration domain $\mathcal{D}(\varepsilon_1) = \left\{\varepsilon_2, \varepsilon_4 \,\vert\, \varepsilon_2 \ge 0, \varepsilon_4 \ge 0, \varepsilon_2 + \varepsilon_4 - \varepsilon_1 \ge 0\right\}$ and the relations $\abs{p_i} = \sqrt{2\,m^i\,\varepsilon_i}$ and $\varepsilon_3 = \varepsilon_2 + \varepsilon_4 - \varepsilon_1$. The (unbounded) domain $\mathcal{D}(\varepsilon_1) \subset \mathbb{R}^2$ simply encodes the physical condition that the individual energies must be non-negative. Note that the $\min$-term in Eq.~\eqref{eq:coll_int_en} expressed by the particle energies reads
\begin{multline}
\label{eq:D_density}
D = \frac{\min(\abs{p_1},\abs{p_2},\abs{p_3},\abs{p_4})}{\abs{p_1}} \\
= \left(\frac{\min\!\left(m^\alpha\varepsilon_1, m^\beta\varepsilon_2, m^\gamma\varepsilon_3, m^\delta\varepsilon_4\right)}{m^\alpha \varepsilon_1}\right)^{1/2}.
\end{multline}

The numerical discretization of the integral \eqref{eq:coll_int_en} should preserve the conservation laws, which result from the interchangeability $\varepsilon_1 \leftrightarrow \varepsilon_3$, $\varepsilon_2 \leftrightarrow \varepsilon_4$ and the pairs $\{\varepsilon_1, \varepsilon_3\} \leftrightarrow \{\varepsilon_2, \varepsilon_4\}$. For this reason, we refrain from using Zakharov transformations \cite{Zakharov1967, SemikozTkachev1997}, and instead opt for a uniform grid for the energy variables, as follows. To adopt the symmetries in the numerical discretization, we first rewrite the integral~\eqref{eq:coll_int_en}:
\begin{equation}
\label{eq:rewrite_coll_int}
\begin{split}
&\hphantom{=} \int_{\mathcal{D}(\varepsilon_1)} \ud\varepsilon_2 \, \ud\varepsilon_4 \, D \left( \mathcal{A}_{\mathrm{quad}}[W]_{1234} + \mathcal{A}_{\mathrm{tr}}[W]_{1234} \right)^{\alpha} \\
&= \int \ud\varepsilon_1' \,\ud\varepsilon_2 \, \ud\varepsilon_3 \, \ud\varepsilon_4 \, \delta(\varepsilon_1' - \varepsilon_2 + \varepsilon_3 - \varepsilon_4) \\
&\qquad \times \delta(\varepsilon_1' - \varepsilon_1) \, D \left( \mathcal{A}_{\mathrm{quad}}[W]_{1234} + \mathcal{A}_{\mathrm{tr}}[W]_{1234} \right)^{\alpha} \\
&= 2 \int_{\mathcal{D}'} \ud s_{13} \,\ud\Delta\varepsilon_{13} \, \ud\Delta\varepsilon_{24} \, \delta(s_{13} + \Delta \varepsilon_{13} - \varepsilon_1) \\
&\quad\qquad \times D \left( \mathcal{A}_{\mathrm{quad}}[W]_{1234} + \mathcal{A}_{\mathrm{tr}}[W]_{1234} \right)^{\alpha},
\end{split}
\end{equation}
where we have used the substitution
\begin{align}
s_{13} &= \frac{1}{2}(\varepsilon_1' + \varepsilon_3), \quad \Delta \varepsilon_{13} = \frac{1}{2}(\varepsilon_1' - \varepsilon_3),\\
s_{24} &= \frac{1}{2}(\varepsilon_2 + \varepsilon_4), \quad \Delta \varepsilon_{24} = \frac{1}{2}(\varepsilon_2 - \varepsilon_4).
\end{align}
The domain $\mathcal{D}'$ of the last integral in~\eqref{eq:rewrite_coll_int} is defined as
\begin{multline}
\mathcal{D}' = \big\{s_{13}, \Delta\varepsilon_{13}, \Delta\varepsilon_{24} \,\big\vert\, s_{13} - \abs{\Delta\varepsilon_{13}} \ge 0, \\
s_{13} - \abs{\Delta\varepsilon_{24}} \ge 0 \big\},
\end{multline}
corresponding to non-negative energies.

Numerically, we store the Wigner matrices $W^{(\alpha)}(\varepsilon_j)$ discretized on a uniform grid for the energy variable:
\begin{equation}
\label{eq:energy_grid}
\varepsilon_j = h \, j, \quad j = 0, 1, \dots, n-1,
\end{equation}
with a small grid spacing $0 < h \ll 1$. The same uniform grid is used to approximate the integration with respect to $s_{13}$, $\Delta\varepsilon_{13}$ and $\Delta\varepsilon_{24}$ in \eqref{eq:rewrite_coll_int}, such that the energy values $\varepsilon_3 = s_{13} - \Delta \varepsilon_{13}$, $\varepsilon_2 = s_{13} + \Delta \varepsilon_{24}$ and $\varepsilon_4 = s_{13} - \Delta \varepsilon_{24}$ are always grid points~\eqref{eq:energy_grid}. Note that $\varepsilon_1 \leftrightarrow \varepsilon_3$ corresponds to $\Delta \varepsilon_{13} \leftrightarrow -\Delta \varepsilon_{13}$ and likewise for $\Delta \varepsilon_{24}$, and that $\{\varepsilon_1, \varepsilon_3\} \leftrightarrow \{\varepsilon_2, \varepsilon_4\}$ corresponds to $\Delta\varepsilon_{13} \leftrightarrow \Delta\varepsilon_{24}$.

Alternative integration schemes (like the apparent Gauss-Laguerre quadrature rule) were also considered but eventually dismissed in favor of the simple trapezoidal rule on a uniform grid. The main advantages are that the conservation laws are respected by the numerical procedure, and that no interpolation of Wigner matrices is required. The uniform discretization has been suggested before \cite{MarkowichPareschi2005}. Unfortunately, the fast algorithm proposed in \cite{MarkowichPareschi2005} cannot simply be used here due to the dependence of $D$ in Eq.~\eqref{eq:D_density} on the particle masses.

Different from the one-dimensional case, a mollification procedure as in~\cite{BoltzmannFermi2012, BoltzmannNonintegrable2013} is not required since the integrals no longer diverge.

Concerning the conservative collision operator $\mathcal{C}_{\mathrm{cons}}$, we perform a change of variables to the energies $\varepsilon_i$ as for the dissipative operator. The integral \eqref{eq:Heff} for the effective Hamiltonian then reads
\begin{multline}
\label{eq:Heff_int}
H_{\mathrm{eff},1}^{\alpha}
= 2\,(2\pi)^2  \, m^\beta m^\gamma m^\delta \\
\times \int \ud\varepsilon_2 \, \ud\varepsilon_3 \, \ud\varepsilon_4 \, \frac{\min(\abs{p_1},\abs{p_2},\abs{p_3},\abs{p_4})}{\abs{p_1}} \\
\times \mathcal{P}\big((\varepsilon_1 - \varepsilon_2 + \varepsilon_3 - \varepsilon_4)^{-1}\big) \, h_{\mathrm{eff}}[W]_{234}^{\alpha}.
\end{multline}
Analytically, the principal value results in the derivative of the integrand, in the sense that
\begin{equation}
\frac{1}{2} \int_{-h}^{h} \ud\varepsilon\, \mathcal{P}\big(\varepsilon^{-1}\big) f(\varepsilon) = h\, f'(0) + \mathcal{O}(h^3)
\end{equation}
for any sufficiently smooth function $f$. In the numerical scheme, we simply omit the grid points for which $\varepsilon_1 - \varepsilon_2 + \varepsilon_3 - \varepsilon_4 = 0$, in order to preserve the conservation laws. The error of this approximation is expected to vanish for grid spacing $h \to 0$.

To solve the Boltzmann equation, we use the explicit midpoint rule for $\mathcal{C} = \mathcal{C}_{\mathrm{diss}} + \mathcal{C}_{\mathrm{cons}}$ as in~\cite{BoltzmannNonintegrable2013}. As advantage, this approach exactly preserves the spin and energy conservation laws.

We have implemented the numerical scheme described so far in plain C code, and use the \emph{MathLink} interface to make the numerical procedures conveniently accessible from Mathematica.

\section{Simulation results}
\label{sec:Simulation}

For the following simulations, we fix an initial Wigner state $W(p,0)$ with particle masses $m^{\ptype{a}} = 1$, $m^{\ptype{b}} = \frac{4}{5}$, $m^{\ptype{c}} = \frac{1}{5}$ and $m^{\ptype{d}} = \frac{1}{2}$. Fig.~\ref{fig:W0} illustrates the $W^{\alpha}(p,0)$ components in dependence of $\abs{p}$. For reference, the analytical formulas of the initial state are recorded in appendix~\ref{sec:InitialW0Analytic}. Note that on the quantum field level in \eqref{eq:interact_diag} and \eqref{eq:interact_cross}, a conservation of masses like $m^{\ptype{a}} + m^{\ptype{c}} = m^{\ptype{b}} + m^{\ptype{d}}$ is not required.

\subsection{Weak interaction: $\beta$ decay}

\begin{figure*}[!ht]
\centering
\subfloat[$m^{\ptype{a}} = 1$]{
\includegraphics[width=0.5\columnwidth]{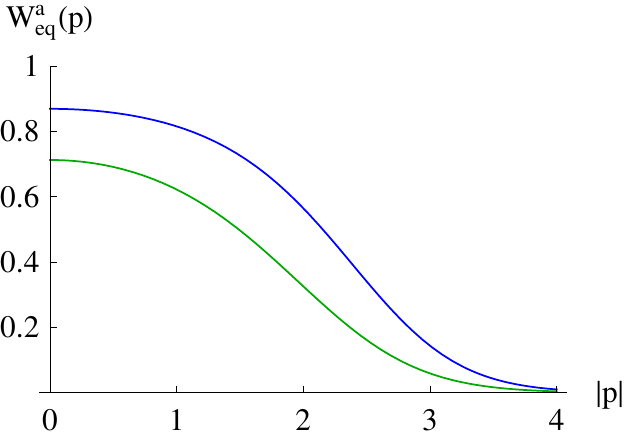}
\label{fig:Weq_n_matrix}}
\subfloat[$m^{\ptype{b}} = 4/5$]{
\includegraphics[width=0.5\columnwidth]{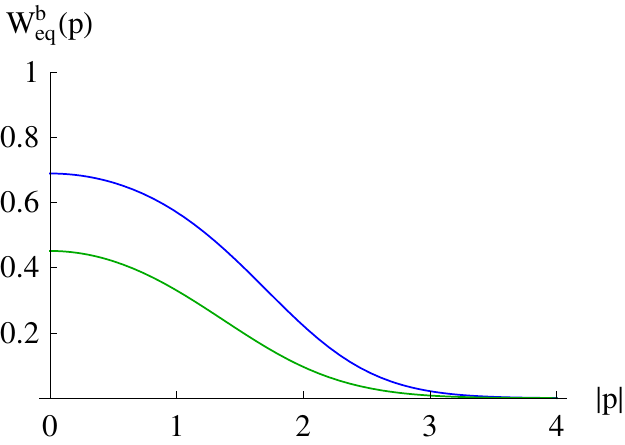}
\label{fig:Weq_p_matrix}}
\subfloat[$m^{\ptype{c}} = 1/5$]{
\includegraphics[width=0.5\columnwidth]{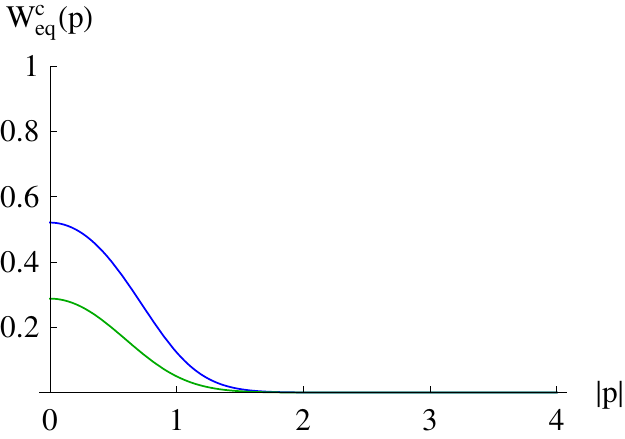}
\label{fig:Weq_nu_matrix}}
\subfloat[$m^{\ptype{d}} = 1/2$]{
\includegraphics[width=0.5\columnwidth]{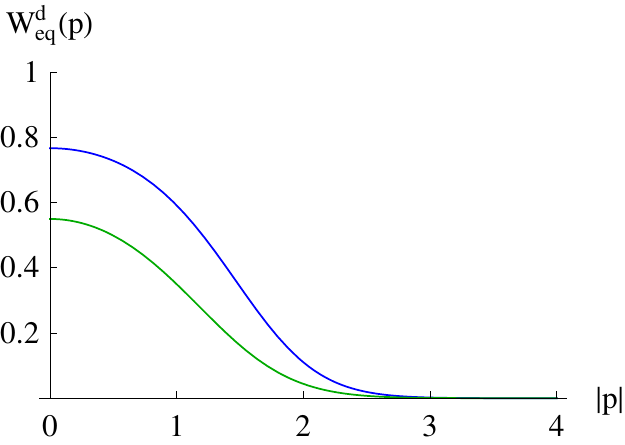}
\label{fig:Weq_e_matrix}}
\caption{(Color online) Diagonal matrix entries of the $t \to \infty$ thermal equilibrium Fermi-Dirac states corresponding to the initial state in Fig.~\ref{fig:W0}, for the case of all $V^{\alpha\beta}$ matrices proportional to the identity matrix. The off-diagonal entries are zero since the states are represented in the eigenbasis of $\rho(t)$, which is conserved in this case. The common inverse temperature $\beta = 0.8193$ and the chemical potentials for each particle type have been determined from the conservation laws.}
\label{fig:Weq_beta}
\end{figure*}

An application of our framework is the $\beta$ decay, i.e., the decay of a neutron $\ptype{n}$ into a proton $\ptype{p}$, an electron $\ptype{e}$ and an antineutrino $\ptype{\overline{\nu}}$. Equivalently, this process can be represented as
\begin{equation}
\label{eq:BetaDecay}
\ptype{n} + \ptype{\nu} \longleftrightarrow \ptype{p} + \ptype{e}.
\end{equation}
The interaction part from Eq.~\eqref{eq:H1defSpatial} is given by
\begin{equation}
H_1 = \frac{4\pi}{g_{\mathrm{w}}^2}\, H_{\beta} ,
\end{equation}
where $g_{\mathrm{w}}$ is the weak coupling factor and
\begin{equation}
\label{eq:H1beta}
\begin{split}
H_{\beta} &= \frac{G_{\mathrm{F}}}{\sqrt{2}} \int \ud^3 x\, \big( \overline{u}^{\ptype{p}} \gamma^{\mu} (C_V + C_A \gamma_5) u^{\ptype{n}} \big) \\
&\qquad \qquad \times \big( \overline{u}^{\ptype{e}} \gamma_{\mu} (1 - \gamma_5) u^{\ptype{\nu}} \big)
\end{split}
\end{equation}
the Hamiltonian of the Fermi theory \cite{Greiner2010}. Einstein summation convention is used for the gamma matrices $\gamma^{\mu}$ and $C_V$, $C_A$ are constants satisfying
\begin{equation}
C_A / C_V = -1.255 \pm 0.006.
\end{equation}
$G_{\mathrm{F}}$ is the Fermi coupling constant. With the relation for the weak coupling constant $\alpha_{\mathrm{w}} = \frac{g_{\mathrm{w}}^2}{4\pi}$, we identify
\begin{equation}
\alpha_{\mathrm{w}} \left( \frac{4\pi}{8 m_{\mathrm{w}}^2} \right) = \frac{g_{\mathrm{w}}^2}{8 m_{\mathrm{w}}^2} = \frac{G_{\mathrm{F}}}{\sqrt{2}},
\end{equation}
where $m_{\mathrm{w}}$ is the mass of the W boson. In our notation of Eq.~\eqref{eq:Hamiltonian} the dimensionless weak coupling
\begin{equation}
\lambda = \alpha_{\mathrm{w}} \approx 0.0339.
\end{equation}
A short calculation shows that the Hamiltonian in Eq.~\eqref{eq:H1beta} can be represented in the form of Eq.~\eqref{eq:H1def} by setting
\begin{align}
\label{eq:Vbeta1}
V^\ptype{n p} &= (C_V - C_A) \mathbbm{1}, & V^\ptype{\nu e} &= \mathbbm{1}, \\
\label{eq:Vbeta2}
V^\ptype{n e} &= \mathbbm{1}            , & V^\ptype{\nu p} &= 2 \, C_A \mathbbm{1}
\end{align}
up to the $g / \sqrt{2}$ prefactor, that is, all interaction matrices are proportional to the identity matrix. Physically, the $\beta$ decay process is independent of spin.

\begin{figure}[!ht]
\centering
\subfloat[time-dependent entropy]{
\includegraphics[width=0.75\columnwidth]{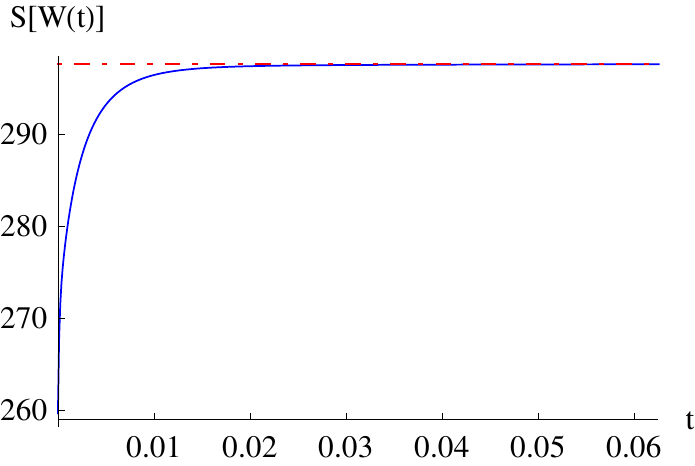}} \\
\subfloat[entropy convergence]{
\includegraphics[width=0.75\columnwidth]{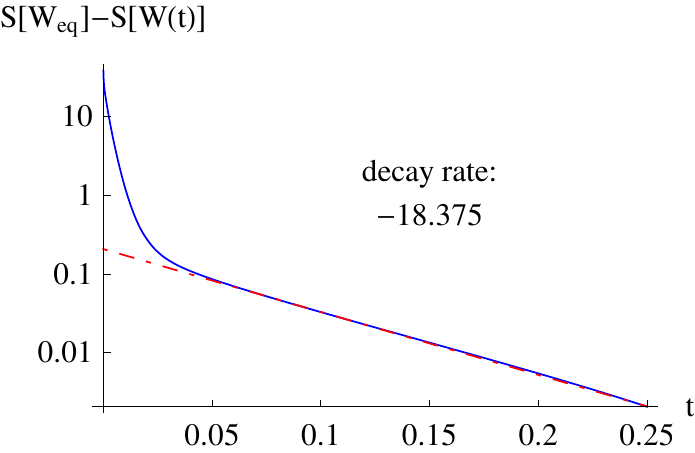}}
\caption{(Color online) Entropy as function of time for the initial state in Fig.~\ref{fig:W0} and the $\beta$ decay interaction matrices in Eqs.~\eqref{eq:Vbeta1} and \eqref{eq:Vbeta2}.}
\label{fig:entropy_beta}
\end{figure}

Fig.~\ref{fig:Weq_beta} illustrates asymptotic thermal Fermi-Dirac equilibrium states as determined from the conservation laws. The equilibrium states are represented in the eigenbasis of the total density $\rho(t)$, which remains constant in time according to Eq.~\eqref{eq:rho_t_constant}. The particle type associations are $\ptype{a}$: neutrons, $\ptype{b}$: protons, $\ptype{c}$: neutrinos and $\ptype{d}$: electrons. The masses are not physically realistic in this model calculation. Our numerical simulation with the interaction matrices in Eqs.~\eqref{eq:Vbeta1} and \eqref{eq:Vbeta2} indeed confirms that the Boltzmann equation drives the initial state in Fig.~\ref{fig:W0} to these thermal equilibrium states. The entropy convergence is visualized in Fig.~\ref{fig:entropy_beta}.

\subsection{Zero outer frame in $\mathcal{V}$}

\begin{figure}[!ht]
\centering
\includegraphics[width=0.75\columnwidth]{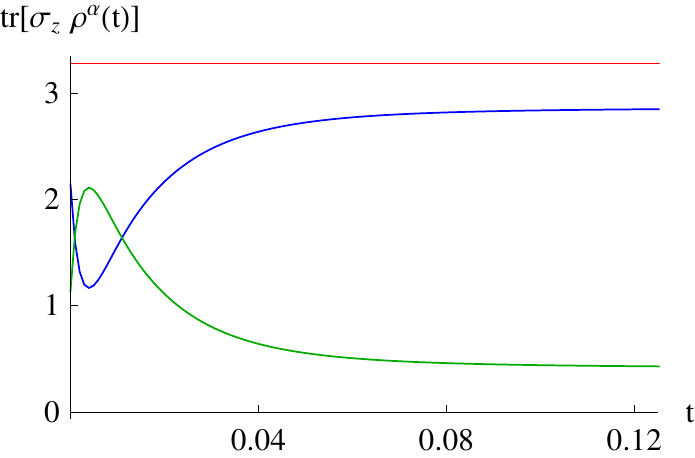}
\caption{(Color online) Projection of the density $\rho^{\alpha}(t)$ onto the $\sigma_z$ Pauli matrix, $\mathrm{tr}[\sigma_z \rho^{\alpha}(t)]$. According to Eq.~\eqref{eq:sigma_z_spin}, the sum over particle types $\ptype{a}$ and $\ptype{c}$ should be conserved (red curve), while the individual types are not necessarily constant in time (blue and green curves for $\ptype{a}$ and $\ptype{c}$, respectively). The curves have been shifted for visual clarity.}
\label{fig:proj_sigma3ac}
\end{figure}

We discuss a simulation with $\mathcal{V}$ matrix (Eq.~\eqref{eq:Vop})
\begin{equation}
\label{eq:Vop_zeroframe_val}
\mathcal{V} = \left(
\begin{array}{cccc}
 0 & 0 & 0 & 0 \\
 0 & -\frac{5}{8} & \frac{1}{3} & 0 \\
 0 & -\frac{1}{4} & \frac{2}{15} & 0 \\
 0 & 0 & 0 & 0 \\
\end{array}
\right)
\end{equation}
The sparsity pattern of $\mathcal{V}$ educes additional conserved quantities, as discussed in Sec.~\ref{sec:Properties}. These conservation laws allow us to predict the asymptotic thermal equilibrium state. Specifically, Fig.~\ref{fig:proj_sigma3ac} shows the projection onto the $\sigma_z$ Pauli matrix: according to Eq.~\eqref{eq:sigma_z_spin}, the sum of types $\ptype{a}$ and $\ptype{c}$ remains constant in time (red curve), but not necessarily the individual types.

\begin{figure}[!ht]
\centering
\subfloat[entropy convergence]{
\includegraphics[width=0.75\columnwidth]{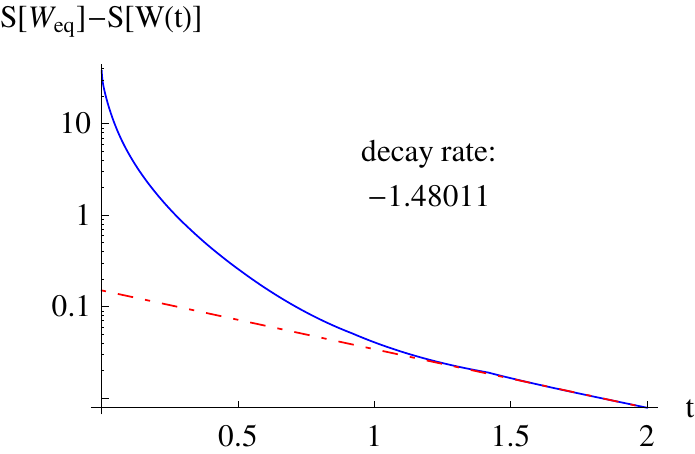}} \\
\subfloat[convergence in $L^1$ norm]{
\includegraphics[width=0.75\columnwidth]{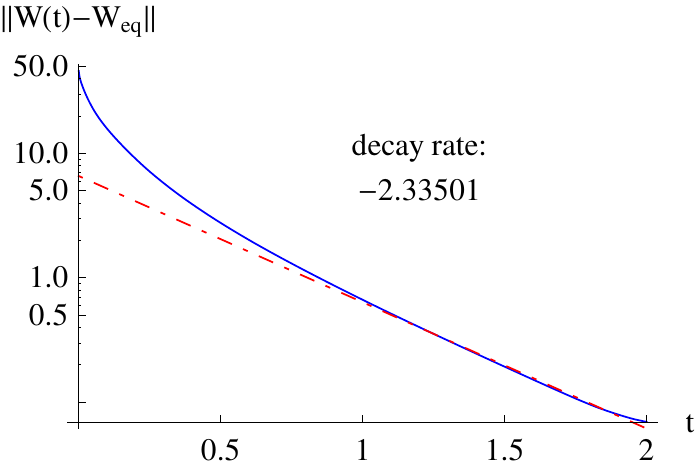}}
\caption{(Color online) Exponential convergence to thermal equilibrium starting from the initial state in Fig.~\ref{fig:W0} and $\mathcal{V}$ matrix in Eq.~\eqref{eq:Vop_zeroframe_val}.}
\label{fig:entropy_HS}
\end{figure}

Fig.~\ref{fig:entropy_HS} illustrates the exponential convergence to thermal equilibrium.

\subsection{Unitary rotation}

\begin{figure}[!ht]
\centering
\includegraphics[width=0.75\columnwidth]{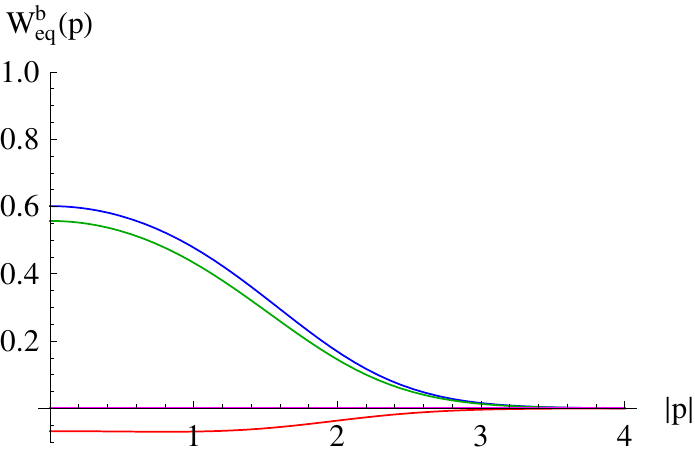}
\caption{(Color online) The asymptotic $t \to \infty$ thermal equilibrium state $W^{\ptype{b}}_{\mathrm{eq}}(p)$ for the $\mathcal{V}$ matrix in Eq.~\eqref{eq:Vop_zeroframe_rot}. The blue and green (upper) curves show the real diagonal entries, and the red curve the real part of the off-diagonal $\ket{\uparrow}\bra{\downarrow}$ entry, respectively. The imaginary part of $\ket{\uparrow}\bra{\downarrow}$ (magenta curve) is zero in this case. The state has non-vanishing off-diagonal entries due to the unitary rotation of the $\ptype{b}$-component in $\mathcal{V}$. The remaining components $W^{\alpha}_{\mathrm{eq}}(p)$ for $\alpha \in \{ \ptype{a}, \ptype{c}, \ptype{d} \}$ are diagonal.}
\label{fig:Wb_zeroframe_rot}
\end{figure}

We transform $\mathcal{V}$ in Eq.~\eqref{eq:Vop_zeroframe_val} by a unitary rotation
\begin{equation}
\mathcal{V} \to \big(U^{\ptype{a}} \otimes U^{\ptype{c}}\big)\, \mathcal{V}\, \big(U^{\ptype{b}} \otimes U^{\ptype{d}}\big)^*
\end{equation}
with $U^{\ptype{a}}$, $U^{\ptype{c}}$ and $U^{\ptype{d}}$ equal to the identity matrix, and
\begin{equation}
U^{\ptype{b}} =
\begin{pmatrix}
\cos(\varphi) & \sin(\varphi) \\
-\sin(\varphi) & \cos(\varphi)
\end{pmatrix} \quad \text{with} \quad \varphi = \pi/5.
\end{equation}
This results in
\begin{equation}
\label{eq:Vop_zeroframe_rot}
\mathcal{V} =
\begin{pmatrix}
0 & 0 & 0 & 0 \\
\frac{1}{3}\sin(\varphi) & -\frac{5}{8} \cos(\varphi) & \frac{1}{3}\cos(\varphi) & \frac{5}{8}\sin(\varphi) \\
 \frac{2}{15}\sin(\varphi) & -\frac{1}{4}\cos(\varphi) & \frac{2}{15}\cos(\varphi) & \frac{1}{4}\sin(\varphi) \\
 0 & 0 & 0 & 0 \\
\end{pmatrix}
\end{equation}
with $\varphi = \pi/5$. The set of conservation laws remains unchanged (``zero outer frame in $\mathcal{V}$ matrix'', last row in table~\ref{tab:ConservedQuantities}) when represented in the basis $\left(U^{\alpha}\right)^* W^{\alpha}(p,t) U^{\alpha}$, although the zero pattern is not evident from Eq.~\eqref{eq:Vop_zeroframe_rot}. Asymptotically, $\left(U^{\alpha}\right)^* W^{\alpha}(p,t) U^{\alpha}$ becomes diagonal for $t \to \infty$, which implies in this case that $W^{\ptype{b}}(p,t)$ will have non-vanishing off-diagonal entries for $t \to \infty$, as visualized in Fig.~\ref{fig:Wb_zeroframe_rot}.

\subsection{Effect of the conservative collision operator}

\begin{figure}[!ht]
\centering
\includegraphics[width=0.75\columnwidth]{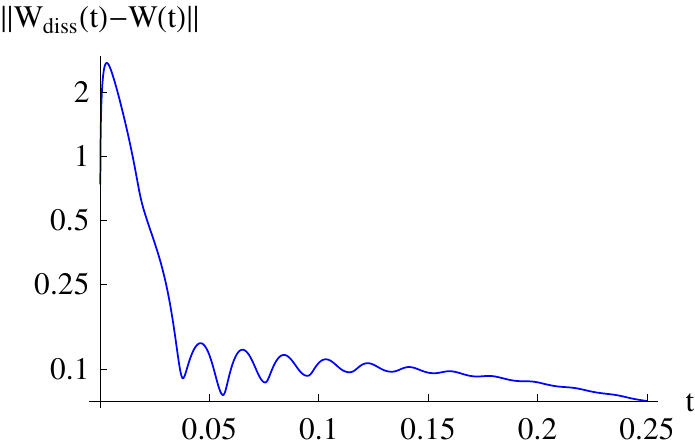}
\caption{(Color online) Distance between $W(t)$ and $W_{\mathrm{diss}}(t)$ obtained from a simulation with $\mathcal{C}_{\mathrm{diss}}$ only, for the $\beta$ decay interaction matrices in Eqs.~\eqref{eq:Vbeta1} and \eqref{eq:Vbeta2}.}
\label{fig:Cd_diff}
\end{figure}

Typically, the conservative collision operator $\mathcal{C}_\mathrm{cons}$ influences the time evolution only slightly. To illustrate this observation quantitatively, we compare a simulation with the physically correct $\mathcal{C} = \mathcal{C}_{\mathrm{diss}} + \mathcal{C}_\mathrm{cons}$ and a simulation using $\mathcal{C}_{\mathrm{diss}}$ only. Fig.~\ref{fig:Cd_diff} shows the corresponding $L^1$ distance between the Wigner states in dependence of time, for the interaction matrices in Eqs.~\eqref{eq:Vbeta1} and \eqref{eq:Vbeta2}. One observes oscillations during the time interval $[0.04, 0.1]$. Note that the distance has to approach zero since the asymptotic ($t \to \infty$) thermal equilibrium state remains the same when omitting $\mathcal{C}_\mathrm{cons}$. In general terms, the trajectories of $W(t)$ are different, but share the same starting point and asymptotic thermal state.

\section{Conclusions and outlook}
\label{sec:Conclusions}

We have disentangled the delicate relationship between the interaction matrices and the time evolution dynamics. As first insight, the interaction matrices $V^{\alpha\beta}$ enter the Boltzmann equation only via the $\mathcal{V}$ matrix defined in Eq.~\eqref{eq:Vop}. Additional conservation laws (table~\ref{tab:ConservedQuantities}) emerge depending on the structure of $\mathcal{V}$. This structure is to be understood modulo unitary rotations of the form \eqref{eq:unitary_invariance}. The conserved quantities in turn determine the asymptotic thermal equilibrium state. Thus, while the particular matrix entries of $\mathcal{V}$ influence the time evolution under the Boltzmann equation, only the structure class of $\mathcal{V}$ dictates the asymptotic state. A complete characterization of all structure classes and corresponding conservation laws is still open, as well as a geometric picture of the manifold of structure classes.

\appendix

\section{Derivation of the multi-component Boltzmann equation}
\label{sec:Derivation}

In this section, we derive the Boltzmann equation starting from the Hamiltonian in Eq.~\eqref{eq:Hamiltonian}. In the spatially homogeneous case, the central quantity is the time-dependent two-point function
\begin{equation}
\label{Wignerdef}
\left\langle\hat{a}_{\sigma}^{\alpha}(p,t)^*\hat{a}_{\tau}^{\beta}\left(p',t\right)\right\rangle = |U|\,\delta(p-p')W_{\sigma\tau}^{\alpha\beta}(p,t),
\end{equation}
which for times up to order $\lambda^{-2}$ will approximately satisfy a kinetic equation. $\langle\cdot\rangle$ denotes the average over the initial state and operators are taken to be in the Heisenberg picture $A(t) = \mathrm{e}^{\mathrm{i} H t}\, A\, \mathrm{e}^{-\mathrm{i} H t}$.
%

\subsection{Basic definitions}
\label{sec:definitions}

Analogous to \cite{DerivationBoltzmann2013}, we introduce spin- and field-dependent vector-valued operators
\begin{equation}
B_{\overline{\mathrm{f}}}(p,t) = \sum_{\substack{\alpha \in \{\ptype{a}, \ptype{b}, \ptype{c}, \ptype{d}\},\\ \sigma \in \{\uparrow, \downarrow\}}} B_{\sigma}^{\alpha}(p,t) \, \overline{\mathrm{f}}^{\alpha}_{\sigma} \, \hat{e}^\alpha_\sigma,
\end{equation}
and 
\begin{equation}
B_\mathrm{g}(p,t) = \sum_{\substack{\alpha \in \{\ptype{a}, \ptype{b}, \ptype{c}, \ptype{d}\},\\ \sigma \in \{\uparrow, \downarrow\}}} B_{\sigma}^{\alpha}(p,t) \, \mathrm{g}^{\alpha}_{\sigma} \, \hat{e}^\alpha_\sigma,
\end{equation}
where $\overline{\mathrm{f}}^\alpha_\sigma$ is the hermitian conjugate of the complex number $\mathrm{f}^\alpha_\sigma$ and $\hat{e}^\alpha_\sigma$ is a unit vector. $B^\alpha_\sigma(p,t)$ is a function in momentum and time. Moreover, we introduce the inner product for spin vectors in two different spin spaces $\overline{\mathrm{f}}$ and $\mathrm{g}$ as
\begin{equation}
B_{\overline{\mathrm{f}}}\odot D_{\mathrm{g}} = \sum_{\alpha,\beta, \sigma,\tau} \overline{\mathrm{f}}^\alpha_\sigma B^\alpha_\sigma\; D^\beta_\tau \mathrm{g}^\beta_\tau \, .
\end{equation}
Thus we will always get a kind of matrix-like term.
A matrix $A$ acts on spin vectors by
\begin{equation}
A\cdot D_{\overline{\mathrm{f}}}=\sum_{\alpha,\beta}\sum_{\sigma,\tau}A^{\alpha\beta}_{\tau\sigma} D^\beta_\sigma \; \overline{\mathrm{f}}^\beta_\sigma \, \hat{e}^\beta_\sigma = (A \cdot D)_{\overline{\mathrm{f}}}.
\end{equation}
Furthermore, we define the term $B^\alpha_{\overline{\mathrm{f}}}\odot C^\beta_\mathrm{g}$ for particle dependent vectors as sum over spins $\sigma$ and $\tau$,
\begin{equation}
B^\alpha_{\overline{\mathrm{f}}}\odot D^\beta_\mathrm{g} = \sum_{\sigma, \tau \in \{\uparrow, \downarrow\}} \overline{\mathrm{f}}^\alpha_\sigma B^\alpha_\sigma\; D^\beta_\tau \mathrm{g}^\beta_\tau \, 
\end{equation}
and spin interaction matrices
\begin{equation}
V^{\alpha\beta}_{\overline{\mathrm{f}}} = \sum_{\sigma, \tau \in \{\uparrow, \downarrow\}} V^{\alpha\beta}_{\sigma\tau}\;\overline{\mathrm{f}}_\tau^\beta
\end{equation}
and
\begin{equation}
V^{\alpha\beta}_\mathrm{g} = \sum_{\sigma, \tau \in \{\uparrow, \downarrow\}} \mathrm{g}_\sigma^\alpha\; V^{\alpha\beta}_{\sigma\tau}.
\end{equation}

\subsection{Time evolution of the two-point correlation function}
\label{sec:timevolutiontwopoint}

Using the introduced notation, we calculate the time evolution of
\begin{multline}
\label{eq:derivativetime}
\frac{\ud }{\ud t}\left\langle\hat{a}_{\overline{\mathrm{f}}}(p_1,t)^*\odot\hat{a}_\mathrm{g}(p_5,t)\right\rangle \\
= \langle\dot{\hat{a}}_{\overline{\mathrm{f}}}(p_1,t)^* \odot\hat{a}_\mathrm{g}(p_5,t)+\hat{a}_{\overline{\mathrm{f}}}(p_1,t)^* \odot \dot{\hat{a}}_\mathrm{g}(p_5,t) \rangle
\end{multline}
with
\begin{equation*}
\hat{a}_{\overline{\mathrm{f}}}(p,t)^* = \sum_{\alpha,\tau}\hat{a}_{\tau}^{\alpha}(p,t)^*\; \overline{\mathrm{f}}^{\alpha}_{\tau}\; \hat{e}^{\alpha}_{\tau}
\end{equation*}
and $a_\mathrm{g}(p_5,t)$ respectively. The dot above a quantity $a(t)$ denotes time derivative: $\frac{\ud}{\ud t} a(t) = \dot{a}(t)$. The time derivative of a field for a single particle type is given by the Heisenberg equation of motion
\begin{equation*}
\begin{split}
\frac{\ud }{\ud t}\hat{a}^\alpha_\tau (p,t)^*
&=\mathrm{i} \left[H,\hat{a}^\alpha_\tau (p,t)^*\right]\\
&= \mathrm{i}\, \mathrm{e}^{-\mathrm{i} H t} \left[H_0 + \lambda H_1,\hat{a}^\alpha_\tau (p)^* \right]\mathrm{e}^{\mathrm{i} H t}.
\end{split}
\end{equation*}
The calculation of the $H_0$ part results in
\begin{equation}
\left[H_0,\hat{a}_{\overline{\mathrm{f}}}(p)^* \right]=\hat{a}_{\overline{\mathrm{f}}}(p)^*\cdot \omega(p).
\end{equation}
Concerning the $H_1$ part, note that the fields depend on different momenta $p_1$ to $p_4$. Specifically for particle type $\ptype{a}$ one obtains
\begin{equation}
\label{eq:commutatorcreator}
\begin{split}
&\left[H_1,\hat{a}^\ptype{a}_\tau(p)^* \right] = -2 \hat{a}^\ptype{a}_\tau (p)^* H_1 + \frac{1}{|U|^{3}} \sum_{p_{1234}}\sum_{\boldsymbol{\sigma}}\delta(\underline{p}) \\
&\quad \big[ \left(\hat{a}^\ptype{b}_{\sigma_1}(p_1)^* V^\ptype{ba}_{\sigma_1 \sigma_2} \delta_{p,p_2}\delta_{\tau,\sigma_2}\right)\left(\hat{a}^\ptype{d}_{\sigma_3} (p_3)^* V^\ptype{dc}_{\sigma_3 \sigma_4}\hat{a}^\ptype{c}_{\sigma_4} (p_4) \right) \\
&\quad + \left(\hat{a}^\ptype{d}_{\sigma_1} (p_1)^* V^\ptype{da}_{\sigma_1 \sigma_2} \delta_{p,p_2}\delta_{\tau,\sigma_2}\right)\left(\hat{a}^\ptype{b}_{\sigma_3} (p_3)^* V^\ptype{bc}_{\sigma_3 \sigma_4}\hat{a}^\ptype{c}_{\sigma_4} (p_4) \right) \big]
\end{split}
\end{equation}
where $\boldsymbol{\sigma} = \{\sigma_1, \sigma_2, \sigma_3, \sigma_4\}$.
%
%
For the following, we define the set
\begin{equation}
\begin{split}
T = \big\{ &(\ptype{a,b,c,d}), (\ptype{a,d,c,b}), (\ptype{b,a,d,c}), (\ptype{b,c,d,a}),\\
&(\ptype{c,d,a,b}), (\ptype{c,b,a,d}), (\ptype{d,c,b,a}), (\ptype{d,a,b,c}) \big\}
\end{split}
\end{equation}
and
\begin{equation}
\label{asterniteration}
\begin{split}
&\mathcal{A}_{\overline{\mathrm{f}}}[h,\mathfrak{a}^*,\mathfrak{a},\mathfrak{a}^*](p_1,t)
=\frac{1}{|U|^{3}}\sum_{p_{234}}\delta(\underline{p}) \sum_{\boldsymbol{\alpha} \in T} h^{\boldsymbol{\alpha}}(t) \\
&\quad \times [\mathfrak{a}^{\alpha_4}(p_4,t)^*\cdot V^{\alpha_4 \alpha_3}\cdot \mathfrak{a}^{\alpha_3}(p_3,t)]\,\mathfrak{a}^{\alpha_2}(p_2,t)^*\cdot V^{\alpha_2 \alpha_1}_{\overline{\mathrm{f}}}
\end{split}
\end{equation}
as well as
\begin{equation}
\label{aiteration}
\begin{split}
&\mathcal{A}_\mathrm{g}[h,\mathfrak{a},\mathfrak{a}^*,\mathfrak{a}](p_1,t) \\
&= \frac{1}{|U|^{3}} \sum_{p_{234}} \delta(\underline{p}) \sum_{\boldsymbol{\alpha}\in T} h^{\boldsymbol{\alpha}}(t)\; V^{\alpha_1 \alpha_2}_\mathrm{g}\cdot\mathfrak{a}^{\alpha_2 }(p_2,t) \\
&\quad \times [\mathfrak{a}^{\alpha_3 }(p_3,t)^*\cdot V^{\alpha_3 \alpha_4} \cdot\mathfrak{a}^{\alpha_4}(p_4,t)].
\end{split}
\end{equation}
Using the invariance under interchanges $p_1 \leftrightarrow p_3$, $p_2 \leftrightarrow p_4$ as well as $(p_1, p_3) \leftrightarrow (p_4, p_2)$, we are able to write the time derivatives of the creation and annihilation operators as
\begin{equation}
\begin{split}
\frac{\ud }{\ud t}\hat{a}_{\overline{\mathrm{f}}}(p_1,t)^{*} &= \mathrm{i}\hat{a}_{\overline{\mathrm{f}}}(p_1,t)^{*} \cdot \omega(p_1) - 2\mathrm{i} \lambda\hat{a}_{\overline{\mathrm{f}}} (p_1,t)^{*} H_{1}(t)\\
&\quad +\mathrm{i}\lambda\mathcal{A}_{\overline{\mathrm{f}}}\left[\mathrm{id},\hat{a}^*,\hat{a},\hat{a}^* \right](p_1,t)
\end{split}
\end{equation}
and
\begin{equation}
\begin{split}
\frac{\ud}{\ud t}\hat{a}_{\mathrm{g}}(p_1,t)=&-\mathrm{i}\omega(p_1) \cdot \hat{a}_{\mathrm{g}}(p_1,t) + 2\mathrm{i} \lambda H_{1}(t) \hat{a}_\mathrm{g}(p_1,t)\\
&-\mathrm{i}\lambda\mathcal{A}_\mathrm{g}\left[\mathrm{id},\hat{a},\hat{a}^*, \hat{a}\right](p_1,t),
\end{split}
\end{equation}
where $\mathrm{id}$ denotes the identity function. In order to simplify calculations, we switch to the interaction picture and define
\begin{equation*}
\mathfrak{a}_{\overline{\mathrm{f}}}(p,t)^* = \hat{a}_{\overline{\mathrm{f}}}(p,t)^*\cdot\mathrm{e}^{-\mathrm{i}\omega(p)t}\mathrm{e}^{\mathrm{i}2\lambda\int^t_0 \ud s\, H_1(s)}
\end{equation*}
and
\begin{equation*}
\mathfrak{a}_{\mathrm{g}}(p,t)=\mathrm{e}^{-\mathrm{i}2\lambda\int^t_0 \ud s\, H_1(s)} \mathrm{e}^{\mathrm{i}\omega(p)t}\cdot\hat{a}_{\mathrm{g}}(p,t),
\end{equation*}
respectively. Thus the dynamics of $\mathfrak{a}(p,t)$ is given by
\begin{equation}
\label{cretime}
\begin{split}
\frac{\ud }{\ud t}\mathfrak{a}_{\overline{\mathrm{f}}}(p_1,t)^{*} = \mathrm{i}\lambda\mathcal{A}_{\overline{\mathrm{f}}}[\mathrm{e}^{-\mathrm{i}\underline{\omega}_{1234}t},\mathfrak{a}^*,\mathfrak{a},\mathfrak{a}^*](p_1,t)
\end{split}
\end{equation}
and
\begin{equation}
\label{antime}
\begin{split}
\frac{\ud}{\ud t}\mathfrak{a}_\mathrm{g}(p,t) = -\mathrm{i}\lambda\mathcal{A}_\mathrm{g}[\mathrm{e}^{\mathrm{i}\underline{\omega}_{1234}t},\mathfrak{a},\mathfrak{a}^*,\mathfrak{a}](p_1,t).
\end{split}
\end{equation}
Moreover,
\begin{equation}
\begin{split}
&\mathcal{A}[\mathrm{e}^{-\mathrm{i}\underline{\omega}_{1234}t},\mathfrak{a}^*,\mathfrak{a},\mathfrak{a}^*](p_1,t) \\
&= \frac{1}{|U|^{3}}\sum_{p_{234}}\delta(\underline{p})\sum_{\boldsymbol{\alpha}\in T}\mathrm{e}^{-\mathrm{i}\omega^{\boldsymbol{\alpha}}_{1234}t} \\
&\quad \times [\mathfrak{a}^{\alpha_4}(p_4,t)^*\cdot V^{\alpha_4 \alpha_3}\cdot \mathfrak{a}^{\alpha_3}(p_3,t)]\,\mathfrak{a}^{\alpha_2}(p_2,t)^*\cdot V^{\alpha_2 \alpha_1}
\end{split}
\end{equation}
and
\begin{equation}
\label{aiteration2}
\begin{split}
&\mathcal{A}[\mathrm{e}^{\mathrm{i}\underline{\omega}_{1234}t},\mathfrak{a},\mathfrak{a}^*,\mathfrak{a}](p_1,t) \\
&=\frac{1}{|U|^{3}}\sum_{p_{234}}\delta(\underline{p})\sum_{\boldsymbol{\alpha}\in T}V^{\alpha_1 \alpha_2}\cdot\mathfrak{a}^{\alpha_2 }(p_2,t) \\
&\quad \times [\mathfrak{a}^{\alpha_3 }(p_3,t)^*\cdot V^{\alpha_3 \alpha_4} \cdot\mathfrak{a}^{\alpha_4}(p_4,t)].
\end{split}
\end{equation}

\subsection{Expansion in powers of $\lambda$}

Iteration of \eqref{cretime} and \eqref{antime} twice up to second order leads to
\begin{multline}
\mathfrak{a}_{\overline{\mathrm{f}}}(p_1,t)^* = \mathfrak{a}_{\overline{\mathrm{f}}} (p_1,0)^* \\
+ \mathrm{i}\lambda\int^t_0 \ud s\, \mathcal{A}_{\overline{\mathrm{f}}}[\mathrm{e}^{-\mathrm{i}\underline{\omega}_{1234}s},\mathfrak{a}^* , \mathfrak{a}, \mathfrak{a}^*] (p_1, s)
\end{multline}
and carrying out the iteration up to order $\lambda^2$ (Duhamel expansion),
\begin{equation}
\begin{split}
&\frac{\ud}{\ud t} \mathfrak{a}_{\overline{\mathrm{f}}}(p_1,t)^* \\
&= \mathrm{i} \lambda\, \mathcal{A}[\mathrm{e}^{-\mathrm{i}\underline{\omega}_{1234}t},\hat{a}^*, \hat{a} ,\hat{a}^*](p_1,0) \\
&\ -\lambda^2 \int^t_0 \ud s \, \mathcal{A}_{\overline{\mathrm{f}}}\big[\mathrm{e}^{-\mathrm{i}\underline{\omega}_{1234}t}, \mathcal{A}[\mathrm{e}^{-\mathrm{i}\underline{\omega}_{4678}s},\hat{a}^*, \hat{a} ,\hat{a}^*], \hat{a} ,\hat{a}^* \big](p_1,s)\\
&\ + \lambda^2 \int^t_0 \ud s \, \mathcal{A}_{\overline{\mathrm{f}}}\big[\mathrm{e}^{-\mathrm{i}\underline{\omega}_{1234}t},\hat{a}^*, \mathcal{A}[\mathrm{e}^{\mathrm{i}\underline{\omega}_{3678}s}, \hat{a} ,\hat{a}^* , \hat{a}] ,\hat{a}^* \big](p_1,s)\\
&\ -\lambda^2 \int^t_0 \ud s \, \mathcal{A}_{\overline{\mathrm{f}}}\big[ \mathrm{e}^{-\mathrm{i}\underline{\omega}_{1234}t},\hat{a}^*, \hat{a}, \mathcal{A}[\mathrm{e}^{-\mathrm{i}\underline{\omega}_{2678}s},\hat{a}^*, \hat{a} ,\hat{a}^*] \big](p_1,s) \\
&\ + \mathcal{O}(\lambda^3) \\
&= \lambda\,\mathfrak{a}^{(1)}_{\overline{\mathrm{f}}}(p,t)^* + \lambda^2\, \mathfrak{a}^{(2)}_{\overline{\mathrm{f}}}(p,t)^* + \mathcal{O}(\lambda^3),
\end{split}
\end{equation} 
where $\mathfrak{a}^{(i)}_{\overline{\mathrm{f}}}(p,t)$ refers to the terms of order $\lambda^i$.

Note that the first term, $\mathfrak{a}^{(0)}_{\overline{\mathrm{f}}}(p,t)^*$, reflects the zero point of the integration and therefore reads
\begin{equation}
\mathfrak{a}^{(0)}_{\overline{\mathrm{f}}}(p,0)^* = \hat{a}_{\overline{\mathrm{f}}}(p)^*.
\end{equation}
Furthermore, the following identity holds
\begin{equation}
\begin{split}
&\langle\mathfrak{a}_{\overline{\mathrm{f}}}(p_1,t)^*\odot \mathfrak{a}_\mathrm{g}(p_5,t)\rangle=\langle\hat{a}_{\overline{\mathrm{f}}}(p_1,t)^* \odot\hat{a}_\mathrm{g}(p_5,t) \rangle.
\end{split}
\end{equation}
Iterating further gives
\begin{multline}
\label{Wtiterated}
\langle\hat{a}_{\overline{\mathrm{f}}}(p_1,t)^*\odot\hat{a}_\mathrm{g}(p_5,t)\rangle = \langle \mathfrak{a}_{\overline{\mathrm{f}}}(p_1,0)\odot \mathfrak{a}_\mathrm{g}(p_5,0) \rangle \\
+\sum^\infty_{n=1} \lambda^n \sum^n_{m=0} \langle \mathfrak{a}_{\overline{\mathrm{f}}}^{(m)}(p_1,t)^* \odot \mathfrak{a}^{(n-m)}_\mathrm{g}(p_5,t) \rangle \\
= \delta(p_1-p_5) \sum_{n=0}^\infty \lambda^n \langle \overline{\mathrm{f}}, W^{(n)}(p_1,t)\cdot \mathrm{g} \rangle,
\end{multline}
where $W^{(n)}(p_1,t)$ is a summation of the relevant terms for $\lambda^n$.
%

\subsubsection{First-order terms}

Starting with the linear $\lambda$ terms, the first thing to do is to calculate $\mathfrak{a}^{(1)}_{\overline{\mathrm{f}}}(p_1,t)$ exactly. Therefore, $\mathfrak{a}^{\alpha}(p,t)$ and $\mathfrak{a}^{\alpha}(p,t)^*$ in \eqref{aiteration} and \eqref{asterniteration} have to be replaced by $\hat{a}^{\alpha}(p)$ and $\hat{a}^{\alpha}(p)^*$. The result is
\begin{equation}
\label{w1def}
\begin{split}
&\delta(p_1-p_5)\langle \overline{\mathrm{f}},W^{(1)}(p_1,t)\cdot \mathrm{g} \rangle \\
&=\langle\mathfrak{a}^{(1)}_{\overline{\mathrm{f}}}(p_1,t)^*\odot\mathfrak{a}^{(0)}_{\mathrm{g}}(p_5,0)+\mathfrak{a}^{(0)}_{\overline{\mathrm{f}}}(p_1,0)^*\odot \mathfrak{a}^{(1)}_{\mathrm{g}}(p_5,t)\rangle\\
&=\mathrm{i}\langle\mathcal{A}_{\overline{\mathrm{f}}}[\mathrm{e}^{-\mathrm{i}\underline{\omega}_{1234}t},\hat{a}^*,\hat{a},\hat{a}^*](p_1,0)^*\odot\hat{a}_\mathrm{g}(p_5)\\
&\ -a_{\overline{\mathrm{f}}}(p_1)^* \odot \mathcal{A}_\mathrm{g} [\mathrm{e}^{\mathrm{i}\underline{\omega}_{1234}t},\hat{a},\hat{a}^*,\hat{a}](p_5,0)\rangle.
\end{split}
\end{equation}
Using Eq.~\eqref{asterniteration} on the first term we get
\begin{equation}
\label{eq:twopointcorrelation10}
\begin{split}
&\langle\mathcal{A}_{\overline{\mathrm{f}}}[\mathrm{e}^{-\mathrm{i}\underline{\omega}_{1234}t},\hat{a}^*,\hat{a},\hat{a}^*](p_1,0)^* \odot \hat{a}_\mathrm{g}(p_5)\rangle \\
&= \mathrm{i} \int^t_0 \ud s\, \frac{1}{|U|^{3}} \sum_{p_{234}} \delta(\underline{p}) \sum_{\boldsymbol{\alpha}\in T} \sum_\beta \mathrm{e}^{-\mathrm{i}\omega^{\boldsymbol{\alpha}}_{1234}s} \\
&\quad \times \Big\langle \left[ \hat{a}^{\alpha_4}(p_4)^*\cdot V^{\alpha_4 \alpha_3}\cdot\hat{a}^{\alpha_3}(p_3)\right] \\
&\qquad \times [(\hat{a}^{\alpha_2}(p_2)^* \cdot V^{\alpha_2 \alpha_1}_{\overline{\mathrm{f}}})\odot\hat{a}^\beta_{\mathrm{g}}(p_5)] \Big\rangle.
\end{split}
\end{equation}

Each summand in $\boldsymbol{\alpha} \in T$ and $\beta$ can be represented by a graph, see Ref.~\cite{DerivationBoltzmann2013}.

Now, to form the average value of Eq.~\eqref{w1def} via Eq.~\eqref{Wignerdef}, we have to perform Wick contractions. If we are averaging over an initial quasi-free state we can partition this average into a product of averages containing only two operators by using the following rule
\begin{equation}
\langle\hat{a}^*_{i_1}\hat{a}_{j_1}\cdot\cdot\cdot\hat{a}^*_{i_n}\hat{a}_{j_n} \rangle = \det\left[ K(i_k,j_l) \right]_{1\leq k,l \leq n},
\end{equation}
where
\begin{equation}
K(i_k,j_l) = \begin{cases}
\langle\hat{a}^*_{i_k}\hat{a}_{j_l}\rangle,  & \text{if}\quad k\leq l, \\
-\langle\hat{a}_{j_l}\hat{a}^*_{i_k}\rangle, & \text{if}\quad k >   l. \\
\end{cases}
\end{equation}
One obtains, for example
\begin{equation}
\label{eq:Wickcontraction10}
\begin{split}
&\langle \hat{a}^\ptype{d}_{\sigma_4}(p_4)^*\hat{a}^\ptype{c}_{\sigma_3}(p_3)\hat{a}^\ptype{b}_{\sigma_2}(p_2)^*\hat{a}^{\beta}_{\tau}(p_5) \rangle = \\
&= \det\!\begin{bmatrix}
\langle\hat{a}^\ptype{d}_{\sigma_4}(p_4)^*\hat{a}^\ptype{c}_{\sigma_3}(p_3)\rangle & \langle\hat{a}^\ptype{d}_{\sigma_4}(p_4)^*\hat{a}^{\beta}_{\tau}(p_5)\rangle\\
-\langle\hat{a}^\ptype{c}_{\sigma_3}(p_3)\hat{a}^\ptype{b}_{\sigma_2}(p_2)^* \rangle& \langle\hat{a}^\ptype{b}_{\sigma_2}(p_2)^*\hat{a}^{\beta}_{\tau}(p_5) \rangle
\end{bmatrix}\\
&= \langle\hat{a}^\ptype{d}_{\sigma_4}(p_4)^*\hat{a}^\ptype{c}_{\sigma_3}(p_3)\rangle \langle\hat{a}^\ptype{b}_{\sigma_2}(p_2)^*\hat{a}^{\beta}_{\tau}(p_5) \rangle\\
&\;\;\;+\langle\hat{a}^\ptype{d}_{\sigma_4}(p_4)^*\hat{a}^{\beta}_{\tau}(p_5)\rangle \langle\hat{a}^\ptype{c}_{\sigma_3}(p_3)\hat{a}^\ptype{b}_{\sigma_2}(p_2)^* \rangle \\
&=0
\end{split}
\end{equation}
for all $\beta \in \{\ptype{a}, \ptype{b}, \ptype{c}, \ptype{d}\}$ since the average value over a pair of annihilator and creator of different particle types is
\begin{equation}
\langle\hat{a}^{\alpha}_{\sigma}(p)^*\hat{a}^{\beta}_{\tau}(p') \rangle=\langle\hat{a}^{\beta}_{\tau}(p')\hat{a}^{\alpha}_{\sigma}(p)^* \rangle=0.
\end{equation}
Similarly all $\lambda$ terms of order one are zero, and therefore Eq.~\eqref{Wtiterated} reduces to
\begin{multline}
\int^t_0 \ud s\frac{\ud }{\ud s}\langle\hat{a}_{\overline{\mathrm{f}}}(p_1,s)^*\odot\hat{a}_\mathrm{g}(p_5,s)\rangle \\
= \lambda^2\delta(p_1-p_5)\int^t_0 \ud s\,\langle \mathrm{f},W^{(2)}(p_1,s)\cdot \mathrm{g}\rangle+\mathcal{O}(\lambda^3).
\end{multline}

\subsubsection{Second-order terms}
\label{sec:lambdaquadrat}

The full $\delta(p_1-p_5) \lambda^2\langle \overline{\mathrm{f}} , W^{(2)}(p_1,t)\cdot \mathrm{g} \rangle$ reads
\begin{equation}
\label{eq:W_second_order}
\begin{split}
&\delta(p_1-p_5) \lambda^2\langle \overline{\mathrm{f}}, W^{(2)}(p_1,t)\cdot \mathrm{g} \rangle \\
&= \int_0^t \ud s\, \langle \mathfrak{a}_{\overline{\mathrm{f}}}^{(1)}(p_1,t)^* \odot \mathfrak{a}_\mathrm{g}^{(1)}(p_5,s) \\
&\quad + \mathfrak{a}_{\overline{\mathrm{f}}}^{(1)}(p_1,s)^* \odot \mathfrak{a}_\mathrm{g}^{(1)}(p_5,t)\rangle + \langle\mathfrak{a}_{\overline{\mathrm{f}}}^{(2)}(p_1,t)^* \odot \hat{a}_\mathrm{g}(p_5) \\
&\quad + \hat{a}_{\overline{\mathrm{f}}}(p_1)^*\odot\mathfrak{a}_\mathrm{g}^{(2)}(p_5,t)\rangle.\\
\end{split}
\end{equation}
Explicitly, the first (1)(1) term is given by
\begin{multline}
\mathfrak{a}_{\overline{\mathrm{f}}}^{(1)}(p_1,t)^*\odot\mathfrak{a}_\mathrm{g}^{(1)}(p_5,s) \\
= \mathcal{A}_{\overline{\mathrm{f}}}[\mathrm{e}^{-\mathrm{i}\underline{\omega}_{1234}t},\hat{a}^*,\hat{a},\hat{a}^*](p_1,t) \\
\odot \mathcal{A}_\mathrm{g}[\mathrm{e}^{\mathrm{i}\underline{\omega}_{1234}s},\hat{a},\hat{a}^*,\hat{a}](p_5,s)
\end{multline}
and the second (1)(1) term results from interchanging $s \leftrightarrow t$. We get
\begin{equation}
\label{eq:11term_a}
\begin{split}
&\mathfrak{a}_{\overline{\mathrm{f}}}^{(1)}(p_1,t)^*\odot\mathfrak{a}_\mathrm{g}^{(1)}(p_5,s)=\\
=& \frac{1}{|U|^{6}}\sum_{p_{234},p_{678}}\delta(\underline{p}_{1234})\delta(\underline{p}_{5678})\sum_{\boldsymbol{\alpha},\boldsymbol{\beta}\,\in T}\mathrm{e}^{-\mathrm{i}\omega^{\boldsymbol{\alpha}}_{1234}t}\mathrm{e}^{\mathrm{i}\omega^{\boldsymbol{\beta}}_{5678}s} \\
&(\hat{a}^{\alpha_4}(p_4)^*\cdot V^{\alpha_4 \alpha_3}\cdot\hat{a}^{\alpha_3}(p_3))(\hat{a}^{\alpha_2}(p_2)^*\cdot V^{\alpha_2 \alpha_1}_{\overline{\mathrm{f}}} )\\
&\odot (V_\mathrm{g}^{\beta_1 \beta_2}\cdot\hat{a}^{\beta_2}(p_6))(\hat{a}^{\beta_3 }(p_7)^*\cdot V^{\beta_3 \beta_4}\cdot\hat{a}^{\beta_4}(p_8)). 
\end{split}
\end{equation}
For what follows, we assume that the initial state $\langle\cdot\rangle$ is quasifree, gauge invariant and invariant under translations. Then the two-point function $\langle a^\alpha_\sigma (p)^* a^\beta_\tau(p') \rangle $ is determined by
\begin{equation}
\langle a^\alpha_\sigma(p)^* a^\beta_\tau(p') \rangle = |U|\delta_{\alpha\beta} \delta(p-p')W^\alpha_{\sigma\tau}(p).
\end{equation}
After taking the average $\langle\cdot\rangle$, the summand on the right of Eq.~\eqref{eq:11term_a} with $\beta_1 = \alpha_1 = \ptype{a}$ is given by
\begin{equation}
\begin{split}
&\frac{1}{|U|^{3}}\sum_{p_{234}} \delta(\underline{p}_{1234}) \delta(\underline{p}_{5234}) \mathrm{e}^{-\mathrm{i}\omega^\ptype{abcd}_{1234}t} \mathrm{e}^{\mathrm{i}\omega^\ptype{abcd}_{5234}s}\\
&\big[\langle\mathrm{f}^{\ptype{a}},V^{\ptype{ab}}\cdot W^{\ptype{b}}_2\cdot V^{\ptype{ba}}\cdot \mathrm{g}^{\ptype{a}} \rangle \, \mathrm{tr}\big[V^{\ptype{cd}}\cdot W^{\ptype{d}}_4 \cdot V^{\ptype{dc}}\cdot\tilde{W}_3^{\ptype{c}} \big] \\
&+\langle\mathrm{f}^{\ptype{a}},V^{\ptype{ad}}\cdot W^{\ptype{d}}_4\cdot V^{\ptype{da}}\cdot \mathrm{g}^{\ptype{a}} \rangle\, \mathrm{tr}\big[ V^{\ptype{cb}}\cdot W^{\ptype{b}}_2 \cdot V^{\ptype{bc}}\cdot\tilde{W}_3^{\ptype{c}} \big] \\
&+\langle\mathrm{f}^{\ptype{a}},V^{\ptype{ab}}\cdot W^{\ptype{b}}_2 \cdot V^{\ptype{bc}}\cdot \tilde{W}_3^{\ptype{c}} \cdot V^{\ptype{cd}}\cdot W^{\ptype{d}}_4 \cdot V^{\ptype{da}}\cdot \mathrm{g}^{\ptype{a}} \rangle\\
&+\langle\mathrm{f}^{\ptype{a}},V^{\ptype{ad}}\cdot W^{\ptype{d}}_4 \cdot V^{\ptype{dc}}\cdot \tilde{W}_3^{\ptype{c}} \cdot V^{\ptype{cb}}\cdot W^{\ptype{b}}_2 \cdot V^{\ptype{ba}} \cdot \mathrm{g}^{\ptype{a}} \rangle\big]\\
\end{split}
\end{equation}
where $\mathrm{f}^{\ptype{a}}$ and $\mathrm{g}^{\ptype{a}}$ are defined as
\begin{equation}
\mathrm{f}^{\ptype{a}} = (\mathrm{f}^{\ptype{a}}_{\uparrow},\mathrm{f}^{\ptype{a}}_{\downarrow})^T, \quad \mathrm{g}^{\ptype{a}} = (\mathrm{g}^{\ptype{a}}_{\uparrow},\mathrm{g}^{\ptype{a}}_{\downarrow})^T.
\end{equation}
The $(\ptype{b}, \ptype{b})$, $(\ptype{c}, \ptype{c})$, and $(\ptype{d}, \ptype{d})$ components are analogous. We obtain the $(\ptype{b}, \ptype{b})$ component by interchanging $\ptype{a} \leftrightarrow \ptype{b}$, $\ptype{c} \leftrightarrow \ptype{d}$, the $(\ptype{c}, \ptype{c})$ component by interchanging $\ptype{a} \leftrightarrow \ptype{c}$, $\ptype{b} \leftrightarrow \ptype{d}$ and the $(\ptype{d}, \ptype{d})$ component by interchanging $\ptype{a} \leftrightarrow \ptype{d}$, $\ptype{b} \leftrightarrow \ptype{c}$.

We collect the components of the Wigner states in a $8\times8$ block-diagonal matrix,
\begin{equation}
\label{eq:Wignergeneral}
W_1=\mathrm{diag}\left[W^{\ptype{a}}_1, W^{\ptype{b}}_1, W^{\ptype{c}}_1, W^{\ptype{d}}_1\right],
\end{equation}
where each entry stands for a $2\times2$-matrix. The interaction potential is summarized by the matrices $V^{\scriptscriptstyle{=}}$ and $V^{\mathsf{x}}$ defined in Eqs.~\eqref{eq:VdiagDef} and \eqref{eq:VcrossDef}, respectively. For the following, define
\begin{equation}
\label{gtraces}
G^{\alpha\beta}_{ij} = \mathrm{tr}\big[V^{\beta\alpha} \cdot W^{\alpha}_j \cdot V^{\alpha\beta}\cdot\tilde{W}^{\beta}_i\big] = (\mathcal{G}^{\alpha\beta}_{ij})^*.
\end{equation}
This definition is used in 
\begin{equation}
\label{eq:tracematrixI}
\mathcal{G}^{\alpha\beta\gamma\delta}_{ij} = \mathrm{diag}\big[G^{\delta\gamma}_{ij}, G^{\gamma\delta}_{ij}, G^{\beta\alpha}_{ij}, G^{\alpha\beta}_{ij}\big] \otimes \mathbbm{1}_{2\times2}
\end{equation}
and
\begin{equation}
\hat{\mathcal{G}}^{\alpha\beta\gamma\delta}_{ij} = \mathrm{diag}\big[G^{\delta\gamma}_{ij}, G^{\alpha\beta}_{ij}, G^{\beta\alpha}_{ij}, G^{\gamma\delta}_{ij}\big]\otimes\mathbbm{1}_{2\times2}
\end{equation}
where the components two and four are exchanged. With these definitions,
\begin{equation}
\label{eq:resulttpc11}
\begin{split}
&\langle \mathfrak{a}_{\overline{\mathrm{f}}}^{(1)}(p_1,t)^* \odot \mathfrak{a}_\mathrm{g}^{(1)}(p_5,s) \rangle = \frac{1}{|U|^{3}} \sum_{p_{234}}\delta(\underline{p}_{1234}) \delta(\underline{p}_{5234})\\
&\times \big\langle\mathrm{f}, \mathrm{e}^{-\mathrm{i}\underline{\omega}_{1234}t} \cdot\mathrm{e}^{\mathrm{i}\underline{\omega}_{5234}s} \\
&\quad \cdot \big(\mathcal{G}^\ptype{abcd}_{34}\cdot V^{\scriptscriptstyle{=}}\cdot W_2\cdot V^{\scriptscriptstyle{=}} + \hat{\mathcal{G}}^\ptype{adcb}_{32}\cdot V^{\mathsf{x}}\cdot W_4\cdot V^{\mathsf{x}} \\
&\quad+ V^{\mathsf{x}}\cdot W_4\cdot V^{\scriptscriptstyle{=}}\cdot \tilde{W}_3\cdot V^{\mathsf{x}}\cdot W_2 \cdot V^{\scriptscriptstyle{=}} \\
&\quad+ V^{\scriptscriptstyle{=}}\cdot W_2\cdot V^{\mathsf{x}}\cdot\tilde{W}_3\cdot V^{\scriptscriptstyle{=}}\cdot W_4 \cdot V^{\mathsf{x}} \big) \cdot\mathrm{g} \big\rangle.
\end{split}
\end{equation}
Furthermore, we define
\begin{equation}
\begin{split}
&\mathcal{D}[W]_{234} \\
&= \mathcal{G}^\ptype{abcd}_{34}\cdot V^{\scriptscriptstyle{=}}\cdot W_2\cdot V^{\scriptscriptstyle{=}}+\hat{\mathcal{G}}^\ptype{adcb}_{32}\cdot V^{\mathsf{x}}\cdot W_4\cdot V^{\mathsf{x}}\\
&\quad + V^{\mathsf{x}}\cdot W_4\cdot V^{\scriptscriptstyle{=}}\cdot\tilde{W}_3\cdot V^{\mathsf{x}}\cdot W_2 \cdot V^{\scriptscriptstyle{=}}\\
&\quad + V^{\scriptscriptstyle{=}}\cdot W_2\cdot V^{\mathsf{x}}\cdot\tilde{W}_3\cdot V^{\scriptscriptstyle{=}}\cdot W_4 \cdot V^{\mathsf{x}}.
\end{split}
\end{equation}
By an analogous calculation,
\begin{equation}
\begin{split}
&\langle \mathfrak{a}_{\overline{\mathrm{f}}}^{(1)}(p_1,s)^* \cdot \mathfrak{a}_\mathrm{g}^{(1)}(p_5,t) \rangle = \frac{1}{|U|^{3}} \sum_{p_2,p_3,p_4} \delta(\underline{p}_{1234}) \delta(\underline{p}_{5234}) \\
&\times \langle\mathrm{f}, \mathrm{e}^{-\mathrm{i}\underline{\omega}_{1234}s} \cdot \mathrm{e}^{\mathrm{i}\underline{\omega}_{5234}t} \cdot \mathcal{D}[W]_{234}\cdot \mathrm{g} \rangle.
\end{split}
\end{equation}
To further simplify the expression, we rearrange the delta functions $\delta(\underline{p}_{1234})\delta(\underline{p}_{5234}) = \delta(p_1-p_5)\delta(\underline{p}_{1234})$ and $p_5$ can be replaced by $p_1$, such that the exponents of the exponential function change signs.

The (2)(0) term is given by
\begin{equation}
\begin{split}
&\mathfrak{a}_{\overline{\mathrm{f}}}^{(2)}(p_1,t)^*\odot\hat{a}_\mathrm{g}(p_5) = - \int_0^t\ud s\\
& \Big( \mathcal{A}_{\overline{\mathrm{f}}}\big[\mathrm{e}^{-\mathrm{i}\underline{\omega}_{1234}t,}, \mathcal{A}[\mathrm{e}^{-\mathrm{i}\underline{\omega}_{4678}s,}, \hat{a}^*, \hat{a}, \hat{a}^*]^*, \hat{a},\hat{a}^*\big](p_1,s)^* \\
& - \mathcal{A}_{\overline{\mathrm{f}}}\big[\mathrm{e}^{-\mathrm{i}\underline{\omega}_{1234}t,}, \hat{a}^*, \mathcal{A}[\mathrm{e}^{\mathrm{i}\underline{\omega}_{3678}s,}, \hat{a}, \hat{a}^*, \hat{a}], \hat{a}^*\big](p_1,s)^*\\
& + \mathcal{A}_{\overline{\mathrm{f}}}[\mathrm{e}^{-\mathrm{i}\underline{\omega}_{1234}t,}, \hat{a}^*, \hat{a}, \mathcal{A}[\mathrm{e}^{-\mathrm{i}\underline{\omega}_{2678}s,}, \hat{a}^*, \hat{a}, \hat{a}^*]^*](p_1,s)^* \Big) \\
&\quad \odot \hat{a}_\mathrm{g}(p_5).
\end{split}
\end{equation}
Thus we get
\begin{equation}
\label{eq:twopoint20}
\begin{split}
&\mathfrak{a}_{\overline{\mathrm{f}}}^{(2)}(p_1,t)^* \odot \hat{a}_\mathrm{g}(p_5) = -\frac{1}{|U|^{6}} \sum_{p_{234},p_{678}} \delta(\underline{p}_{1234}) \\
&\times \sum_\beta\sum_{\boldsymbol{\alpha} \in T} \int_0^t \ud s\, \mathrm{e}^{-\mathrm{i}\omega_{1234}^{\boldsymbol{\alpha}}t} ( X - Y + Z ) \odot \hat{a}^{\beta}_\mathrm{g}(p_5)
\end{split}
\end{equation}
with
\begin{equation}
\begin{split}
&X = \sum_{(\alpha_2,\beta_2,\beta_3\beta_4) \in T} \delta(\underline{p}_{2678}) \ \mathrm{e}^{-\mathrm{i}\omega_{2678}^{\alpha_2\beta_2\beta_3\beta_4}s} \\
& \big(\hat{a}^{\alpha_4}(p_4)^*\cdot V^{\alpha_4 \alpha_3} \cdot\hat{a}^{\alpha_3}(p_3)\big) \big(\hat{a}^{\beta_4}(p_8)^*\cdot V^{\beta_4 \beta_3} \cdot \hat{a}^{\beta_3}(p_7)\big)\\
& \big(\hat{a}^{\beta_2}(p_6)^*\cdot V^{\beta_2 \alpha_2} \big)\, V^{\alpha_2 \alpha_1}_{\overline{\mathrm{f}}}
\end{split}
\end{equation}
and
\begin{equation}
\begin{split}
&Y = \sum_{(\alpha_3,\beta_2,\beta_3\beta_4) \in T} \delta(\underline{p}_{3678}) \ \mathrm{e}^{\mathrm{i}\omega_{3678}^{\alpha_3\beta_2\beta_3\beta_4}s} \\
& \big(\hat{a}^{\alpha_4}_{\sigma_{\alpha_4}}(p_4)^*\cdot V^{\alpha_4 \alpha_3} \cdot V^{\alpha_3 \beta_2}\cdot\hat{a}^{\beta_2}(p_6)\big) \\
& \big(\hat{a}^{\beta_3}(p_7)^*\cdot V^{\beta_3 \beta_4}\cdot\hat{a}^{\beta_4}(p_8)\big)\;\hat{a}^{\alpha_2}(p_2)^*\cdot V^{\alpha_2 \alpha_1}_{\overline{\mathrm{f}}} \\
\end{split}
\end{equation}
and
\begin{equation}
\begin{split}
&Z = \sum_{(\alpha_4,\beta_2,\beta_3\beta_4) \in T} \delta(\underline{p}_{4678})\ \mathrm{e}^{-\mathrm{i}\omega_{4678}^{\alpha_4\beta_2\beta_3\beta_4}s} \\
& \big(\hat{a}^{\beta_4}(p_8)^*\cdot V^{\beta_4 \beta_3}\cdot\hat{a}^{\beta_3}(p_7)\big) \\
& \big(\hat{a}^{\beta_2}(p_6)^* \cdot V^{\beta_2 \alpha_4} \cdot V^{\alpha_4 \alpha_3} \cdot \hat{a}^{\alpha_3}(p_3)\big) \cdot \hat{a}^{\alpha_2}(p_2)^*\ V^{\alpha_2 \alpha_1}_{\overline{\mathrm{f}}}.
\end{split}
\end{equation}
Only terms with complementary creation and annihilation operators of the same particle type are non-zero when taking the average. For the following, we introduce
\begin{equation}
F_{i}^{\alpha\beta} = \mathrm{tr}\big[V^{\beta\alpha}\cdot W^{\alpha}_i\cdot V^{\alpha\beta}\big] = (\mathcal{F}_{i}^{\alpha\beta})^*
\end{equation}
which is summarized by the block-diagonal matrices
\begin{equation}
\label{eq:singletrace}
\mathcal{F}^{\alpha\beta\gamma\delta}_{i} = \mathrm{diag}\big[F^{\delta\gamma}_{i},F^{\gamma\delta}_{i},F^{\beta\alpha}_{i},F^{\alpha\beta}_{i}\big]\otimes\mathbbm{1}_{2\times2}
\end{equation}
and
\begin{equation}
\label{eq:singletracematrix}
\hat{\mathcal{F}}^{\alpha\beta\gamma\delta}_{i} = \mathrm{diag}\big[F^{\delta\gamma}_{i},F^{\alpha\beta}_{i},F^{\beta\alpha}_{i},F^{\gamma\delta}_{i}\big]\otimes\mathbbm{1}_{2\times2}.
\end{equation}
Note that the second and fourth entry on the right in \eqref{eq:singletracematrix} are exchanged as compared to \eqref{eq:singletrace}. As heuristic motivation, the calculation for the $H^\ptype{abcd}_1$-part in the Hamiltonian is analogous to the $H^\ptype{adcb}_1$-part with particles $\ptype{b}$ and $\ptype{d}$ exchanged. In summary, one obtains
\begin{equation}
\label{eq:tpf20retult}
\begin{split}
&\langle\mathfrak{a}_{\overline{\mathrm{f}}}^{(2)}(p_1,t)^* \odot \hat{a}_\mathrm{g}(p_5) \rangle = \delta(p_1 - p_5) \frac{1}{|U|^{3}} \sum_{p_{234}} \delta(\underline{p}_{1234}) \\
&\quad \times \int^t_0 \ud s\, \big\langle\mathrm{f}, \mathrm{e}^{-\mathrm{i}\underline{\omega}_{1234}(t-s)} \cdot \mathcal{B}[W]_{1234} \cdot \mathrm{g} \big\rangle
\end{split}
\end{equation}
with the definition
\begin{equation}
\begin{split}
&\mathcal{B}[W]_{1234} = \\
&-\big( \mathcal{G}^\ptype{abcd}_{34}\cdot V^{\scriptscriptstyle{=}}\cdot V^{\scriptscriptstyle{=}}\cdot W_5 + \hat{\mathcal{G}}^\ptype{adcb}_{32}\cdot V^{\mathsf{x}}\cdot V^{\mathsf{x}}\cdot W_5 \\
&+V^{\scriptscriptstyle{=}}\cdot V^{\mathsf{x}}\cdot \tilde{W}_3\cdot V^{\scriptscriptstyle{=}}\cdot W_4 \cdot V^{\mathsf{x}}\cdot W_5 \\
&+ V^{\mathsf{x}}\cdot V^{\scriptscriptstyle{=}}\cdot \tilde{W}_3 \cdot V^{\mathsf{x}}\cdot W_2 \cdot V^{\scriptscriptstyle{=}}\cdot W_5 \big) \\
&+\big( \mathcal{F}^\ptype{abcd}_{4}\cdot V^{\scriptscriptstyle{=}}\cdot \tilde{W}_2\cdot V^{\scriptscriptstyle{=}}\cdot W_5 \\
&+ \hat{\mathcal{F}}^\ptype{adcb}_{2}\cdot V^{\mathsf{x}}\cdot \tilde{W}_4\cdot V^{\mathsf{x}}\cdot W_5\\
&+V^{\scriptscriptstyle{=}}\cdot\tilde{W}_2\cdot V^{\mathsf{x}}\cdot V^{\scriptscriptstyle{=}}\cdot W_4\cdot V^{\mathsf{x}}\cdot W_5 \\
&+ V^{\mathsf{x}}\cdot\tilde{W}_4\cdot V^{\scriptscriptstyle{=}}\cdot V^{\mathsf{x}}\cdot W_2\cdot V^{\scriptscriptstyle{=}}\cdot W_5 \big)\\
&-\big( \mathcal{F}^\ptype{badc}_{3}\cdot V^{\scriptscriptstyle{=}}\cdot \tilde{W}_2\cdot V^{\scriptscriptstyle{=}}\cdot W_5 \\
&+ \hat{\mathcal{F}}^\ptype{dabc}_{3}\cdot V^{\mathsf{x}}\cdot \tilde{W}_4\cdot V^{\mathsf{x}}\cdot W_5\\
&+V^{\scriptscriptstyle{=}}\cdot\tilde{W}_2\cdot V^{\mathsf{x}}\cdot W_3\cdot V^{\scriptscriptstyle{=}}\cdot V^{\mathsf{x}}\cdot W_5 \\
&+ V^{\mathsf{x}}\cdot\tilde{W}_4\cdot V^{\scriptscriptstyle{=}}\cdot W_3\cdot V^{\mathsf{x}}\cdot V^{\scriptscriptstyle{=}}\cdot W_5 \big).
\end{split}
\end{equation}
For the (0)(2) term of Eq.~\eqref{eq:W_second_order} we get analogously
\begin{equation}
\label{eq:tpf02result}
\begin{split}
&\langle\hat{a}_{\overline{\mathrm{f}}}(p_1)^*\odot\mathfrak{a}_\mathrm{g}^{(2)}(p_5,t) \rangle \\
& = \delta(p_1 - p_5) \frac{1}{|U|^{3}}\sum_{p_{234}} \delta(\underline{p}_{1234}) \\
& \quad \times \int^s_0 \ud s\, \big\langle \mathrm{f}, \mathrm{e}^{i\underline{\omega}_{5234}(t-s)}\cdot\mathcal{B}[W]_{1234}^* \big\rangle.
\end{split}
\end{equation}

\subsection{The limit $\lambda \to 0$, $t = \mathcal{O}(\lambda^{-2})$}

We take the infinite volume limit $\ell \rightarrow \infty$ of $U = U_\ell = [-\ell,\ell]^{\ptype{d}}$ and subsequently the kinetic limit $\lambda \rightarrow 0$ together with rescaling $t \rightarrow \lambda^{-2}t$. Defining
\begin{equation}
\begin{split}
&H(p_{1234},t) = \int_0^t\ud s\, \int_0^s\ud s'\, \delta(\underline{p}_{1234}) \\
&\ \times \big(\mathrm{e}^{\mathrm{i}\underline{\omega}_{1234}(s-s')} \cdot \left(\mathcal{D}[W]_{234} + \mathcal{B}[W]_{1234}^* \right) \\
&\quad + \mathrm{e}^{-\mathrm{i}\underline{\omega}_{1234}(s-s')}\cdot \left(\mathcal{D}[W]_{234} + \mathcal{B}[W]_{1234} \right) \big),
\end{split}
\end{equation}
we get
\begin{equation}
\label{eq:volumelimit}
\begin{split}
&\delta(p_1-p_5) \, \langle \mathrm{f}, W^{(2)}(p_1,t)\cdot \mathrm{g}\rangle \\
&= \int^t_0 \ud s \frac{\ud }{\ud s} \sum_{m=0}^2 \langle\hat{a}^*_{\overline{\mathrm{f}}}(p_1,s)^{(m)} \odot \hat{a}_\mathrm{g}(p_5,s)^{(2-m)} \rangle \\
& = \delta(p_1-p_5) \, \frac{1}{|U_{\ell}|^3}\,\sum_{p_{234} \in \hat{U}^3} \langle \mathrm{f}, H(p_{1234},t)\cdot \mathrm{g}\rangle.
\end{split}
\end{equation}
In the limit ${\ell\rightarrow \infty}$ we obtain the Riemann integral
\begin{equation}
\begin{split}
&\lim_{\ell \to \infty} \frac{1}{|U_{\ell}|} \sum_{p \in \frac{\pi}{\ell} \mathbb{Z}^d } f(p) = \frac{1}{2\pi} \int_{\mathbb{R}^d} \ud p\, f(p).
\end{split}
\end{equation}
Thus
\begin{equation}
\begin{split}
W^{(2)}(p_1,t) &= \lim_{\ell \rightarrow \infty} \frac{1}{|U_{\ell}|^3}\,\sum_{p_{234} \in \frac{\pi}{\ell} \mathbb{Z}^{3d} } H(p_{1234},t) \\
&= \frac{1}{(2\pi)^3} \int_{\mathbb{R}^{3d}} \ud p_{234} \, H(p_{1234},t) .
\end{split}
\end{equation}

The collision operator is determined by taking at second order the limit $\lambda \rightarrow 0$ and simultaneously long times $\lambda^{-2} t$ with $t$ of order $1$. More explicitly,
\begin{equation}
t\,\mathcal{C}[W](p) = \lim_{\lambda \to 0} \lambda^2\, W^{(2)}\big(p,\lambda^{-2} t\big),
\end{equation}
To evaluate the limit, we use
\begin{equation}
\begin{split}
&\lim_{\lambda \rightarrow 0} \lambda^2 \int^{\lambda^{-2} t}_0 \ud s\, \int^{s}_0\ud s'\, \mathrm{e}^{\pm\mathrm{i}\omega_{1234}(s-s')} \\
&= t \int^\infty_0 \ud s\, \mathrm{e}^{\pm\mathrm{i}\omega_{1234}s} = t \left(\pm \mathrm{i} \, \mathcal{P}\big(\omega_{1234}^{-1}\big) + \pi\, \delta(\omega_{1234}) \right)
\end{split}
\end{equation}
where $\mathcal{P}$ denotes the principal value integral. Thus
\begin{equation}
\begin{split}
&\lim_{\lambda \rightarrow 0}\; \lambda^2\, W^{(2)}(p_1, \lambda^{-2} t) \\
&= \frac{t \, \pi}{(2\pi)^3} \int_{\mathbb{R}^{3d}}\ud p_{234}\, \delta(\underline{p}) \delta(\underline{\omega}_{1234}) \\
&\hspace{50pt} \cdot \left(2\,\mathcal{D}[W]_{234} +\mathcal{B}[W]_{1234}^* + \mathcal{B}[W]_{1234}\right) \\
& + \frac{t\, \mathrm{i}}{(2\pi)^3} \int_{\mathbb{R}^{3d}}\ud p_{234}\, \delta(\underline{p}) \mathcal{P}\left(\underline{\omega}^{-1}_{1234}\right) \\
&\hspace{50pt} \cdot \left(\mathcal{B}[W]_{1234}^* - \mathcal{B}[W]_{1234}\right)
\end{split}
\end{equation}
where $\mathcal{P}\left(\underline{\omega}^{-1}_{1234}\right)$ must be considered as principal value applied to every component and similarly $\delta(\underline{\omega}_{1234})$ as a matrix of delta functions
\begin{equation}
\begin{split}
\mathcal{P}\big(\underline{\omega}^{-1}_{1234}\big)
&= \mathrm{diag}\Big[\mathcal{P}\Big(\frac{1}{\omega_{1234}^\ptype{abcd}}\Big), \mathcal{P}\Big(\frac{1}{-\omega_{2143}^\ptype{abcd}}\Big), \\
&\quad \mathcal{P}\Big(\frac{1}{\omega_{3412}^\ptype{abcd}}\Big), \mathcal{P}\Big(\frac{1}{-\omega_{4321}^\ptype{abcd}}\Big)\Big] \otimes \mathbbm{1}_{2\times2}
\end{split}
\end{equation}
and
\begin{equation}
\begin{split}
\delta(\underline{\omega}_{1234})
&= \mathrm{diag}\Big[ \delta\big(\omega_{1234}^\ptype{abcd}\big), \delta\big(-\omega_{2143}^\ptype{abcd}\big), \\
&\hspace{35pt} \delta\big(\omega_{3412}^\ptype{abcd}\big), \delta\big(-\omega_{4321}^\ptype{abcd}\big) \Big]\otimes\mathbbm{1}_{2\times2}.
\end{split}
\end{equation}
We obtain
\begin{equation}
\frac{\partial}{\partial t}W(p_1,t) = \mathcal{C}_{\mathrm{diss}}[W](p_1,t) + \mathcal{C}_{\mathrm{cons}}[W](p_1,t)
\end{equation}
with
\begin{multline}
\mathcal{C}_{\mathrm{diss}}[W](p_1,t) = \frac{\pi}{(2\pi)^3} \int_{\mathbb{R}^{3d}}\ud p_{234}\, \delta(\underline{p}) \delta(\underline{\omega}_{1234}) \\
\quad\cdot \left(2\,\mathcal{D}[W]_{234} +\mathcal{B}[W]_{1234} + \mathcal{B}[W]_{1234}^* \right)
\end{multline}
and
\begin{multline}
\mathcal{C}_{\mathrm{cons}}[W](p_1,t) = \frac{\mathrm{i}}{(2\pi)^3} \int_{\mathbb{R}^{3d}}\ud p_{234}\, \delta(\underline{p}) \mathcal{P} \left(\underline{\omega}^{-1}_{1234}\right) \\
\cdot \left(\mathcal{B}[W]_{1234}^* - \mathcal{B}[W]_{1234}\right).
\end{multline}
Note that without spin interaction the conservative part would vanish since $\mathcal{B}$ and $\mathcal{B}^*$ cancel out. Finally, algebraic reformulation and using the symmetry properties leads to
\begin{multline}
\mathcal{A}_{\mathrm{quad}}[W]_{1234} + \mathcal{A}_{\mathrm{tr}}[W]_{1234} \\
= 2\,\mathcal{D}[W]_{234} +\mathcal{B}[W]_{1234} + \mathcal{B}[W]_{1234}^*
\end{multline}
and
\begin{equation}
\big[ h_{\mathrm{eff}}[W]_{234}, W_1 \big] = \mathcal{B}[W]_{1234}^* - \mathcal{B}[W]_{1234}.
\end{equation}

\newpage

\section{Initial Wigner state $W(p,0)$}
\label{sec:InitialW0Analytic}

For reproducibility, we record the analytical formula of the initial Wigner state $W(p,0)$ used in the simulations (Fig.~\ref{fig:W0}). We specify the state in dependence of the energy $\varepsilon$, which is related to the momentum via the dispersion relation $\varepsilon = \omega^{\alpha}(p)$ for particle type $\alpha$, see Eq.~\eqref{eq:omegaAlphaDef}.

The $\ptype{a}$-component is
\begin{equation}
\begin{split}
W^{\ptype{a}}_{\uparrow\uparrow}(\varepsilon,0) &= \tfrac{5}{2}\, \mathrm{e}^{-2\varepsilon} \left(\varepsilon^2 + \tfrac{1}{4}\right)^2 , \\
W^{\ptype{a}}_{\uparrow\downarrow}(\varepsilon,0) &= 42\, \mathrm{e}^{2\mathrm{i} (\varepsilon-1/3) - \frac{1}{2}(\varepsilon-15/4)^2 - 2\varepsilon} , \\
W^{\ptype{a}}_{\downarrow\downarrow}(\varepsilon,0) &= \tfrac{1}{6}\, \mathrm{erfc}(\varepsilon-6)\, \mathrm{e}^{-2\varepsilon/3}\, \mathrm{atan}(\varepsilon+1), \\
&\quad \times \left(2\,\mathrm{erf}\big(\tfrac{\varepsilon}{2}\big) + \tfrac{1}{8}\right) \left(2 + \tfrac{1}{2} \sin(3\,\varepsilon)\right) ,
\end{split}
\end{equation}
the $\ptype{b}$-component reads
\begin{equation}
\begin{split}
W^{\ptype{b}}_{\uparrow\uparrow}(\varepsilon,0) &= \tfrac{2}{3}\, \left(2+\sin(2\,\varepsilon)\right) \left(2 + \Gamma(1+\varepsilon)\right)^{-1} , \\
W^{\ptype{b}}_{\uparrow\downarrow}(\varepsilon,0) &= \tfrac{1}{2}\, \zeta\!\left((1+\tfrac{\mathrm{i}}{2})\,\varepsilon\right) \mathrm{e}^{-2\varepsilon} , \\
W^{\ptype{b}}_{\downarrow\downarrow}(\varepsilon,0) &= \mathrm{e}^{-(1+2\varepsilon/3)} ,
\end{split}
\end{equation}

\newpage

the $\ptype{c}$-component
\begin{equation}
\begin{split}
W^{\ptype{c}}_{\uparrow\uparrow}(\varepsilon,0) &= \tfrac{2}{3}\, \mathrm{erfc}(\tfrac{\varepsilon}{2}) \left(\varepsilon^2 + \tfrac{4}{5}\right) \left( \tfrac{3}{5} + \tfrac{1}{6} \varepsilon^2 \right), \\
W^{\ptype{c}}_{\uparrow\downarrow}(\varepsilon,0) &= \tfrac{1}{2}\, \mathrm{e}^{-3\varepsilon/2} \left(1+\mathrm{erf}(\varepsilon-2)\right) \mathrm{erfc}(\varepsilon-6) \\
&\quad \times \left( \tfrac{2}{5} - \mathrm{i}\,\varepsilon + 4 (1+\mathrm{i})\, \varepsilon \sin(\varepsilon)^2 \right), \\
W^{\ptype{c}}_{\downarrow\downarrow}(\varepsilon,0) &= \mathrm{erfc}(\varepsilon-6) \, \mathrm{e}^{-\varepsilon/2} \left(1 + \sin(\varepsilon)^2\right) \left(3 + \tfrac{3}{5} \varepsilon \right)^{-1} ,
\end{split}
\end{equation}
and the $\ptype{d}$-component
\begin{equation}
\begin{split}
W^{\ptype{d}}_{\uparrow\uparrow}(\varepsilon,0) &= \tfrac{3}{4\pi}\, \mathrm{erfc}(\varepsilon-7)\, \mathrm{e}^{-\varepsilon/2}\, \mathrm{Si}\!\left( 6\,\varepsilon + \tfrac{1}{2} \right), \\
W^{\ptype{d}}_{\uparrow\downarrow}(\varepsilon,0) &= \tfrac{1}{24}\, \mathrm{erfc}(\varepsilon-6)\, \mathrm{e}^{\mathrm{i} \pi 6/7 - 3\varepsilon/2} \\
&\quad \times \sqrt{\varepsilon} \left( 15-18\varepsilon + 3\varepsilon^2 \right), \\
W^{\ptype{d}}_{\downarrow\downarrow}(\varepsilon,0) &= \mathrm{Ai}(\varepsilon-1) .
\end{split}
\end{equation}
Here $\zeta(s)$ is the Riemann zeta function, $\mathrm{erf}(z)$ the error function, $\mathrm{erfc}(z)$ the complementary error function, $\mathrm{Si}(z)$ the sine integral function and $\mathrm{Ai}(x)$ the Airy function.

The off-diagonal entries $W^{\alpha}_{\downarrow\uparrow}(\varepsilon,0)$ are respective complex conjugates of $W^{\alpha}_{\uparrow\downarrow}(\varepsilon,0)$ since $W(\varepsilon,0)$ is Hermitian.


\end{document}